\batchmode
\makeatletter
\def\input@path{{"C:/Trabajo laptop/Mis articulos/Finished/Coexisting point-extended PMBM/Accepted/"}}
\makeatother
\documentclass[10pt]{IEEEtran}
\usepackage[latin9]{inputenc}
\usepackage{float}
\usepackage{amsmath}
\usepackage{amsthm}
\usepackage{amssymb}
\usepackage{graphicx}
\PassOptionsToPackage{normalem}{ulem}
\usepackage{ulem}

\makeatletter

\providecommand{\tabularnewline}{\\}
\floatstyle{ruled}
\newfloat{algorithm}{tbp}{loa}
\providecommand{\algorithmname}{Algorithm}
\floatname{algorithm}{\protect\algorithmname}

\theoremstyle{plain}
\newtheorem{thm}{\protect\theoremname}
\theoremstyle{plain}
\newtheorem{lem}[thm]{\protect\lemmaname}

\usepackage{cite}
\allowdisplaybreaks 

\usepackage[margin=8pt,font=footnotesize]{caption}
\usepackage{algorithm}
\usepackage{algpseudocode}
\usepackage{amsmath}  

\makeatother

\providecommand{\lemmaname}{Lemma}
\providecommand{\theoremname}{Theorem}

\begin{document}
\title{A Poisson multi-Bernoulli mixture filter for coexisting point and
extended targets}
\author{Ángel F. García-Fernández, Jason L. Williams, Lennart Svensson, Yuxuan
Xia\thanks{A. F. García-Fernández is with the Department of Electrical Engineering and Electronics, University of Liverpool, Liverpool L69 3GJ, United Kingdom (angel.garcia-fernandez@liverpool.ac.uk). He is also with  the ARIES Research Centre, Universidad Antonio de Nebrija,  Madrid, Spain. J. L. Williams is with the Commonwealth Scientific and Industrial Research Organization (jason.williams@data61.csiro.au). L. Svensson and Y. Xia are with the Department of Electrical Engineering, Chalmers University of Technology, SE-412 96 Gothenburg, Sweden (firstname.lastname@chalmers.se).} }
\maketitle
\begin{abstract}
This paper proposes a Poisson multi-Bernoulli mixture (PMBM) filter
for coexisting point and extended targets, i.e., for scenarios where
there may be simultaneous point and extended targets. The PMBM filter
provides a recursion to compute the multi-target filtering posterior
based on probabilistic information on data associations, and single-target
predictions and updates. In this paper, we first derive the PMBM filter
update for a generalised measurement model, which can include measurements
originated from point and extended targets. Second, we propose a single-target
space that accommodates both point and extended targets and derive
the filtering recursion that propagates Gaussian densities for point
targets and gamma Gaussian inverse Wishart densities for extended
targets. As a computationally efficient approximation of the PMBM
filter, we also develop a Poisson multi-Bernoulli (PMB) filter for
coexisting point and extended targets. The resulting filters are analysed
via numerical simulations.
\end{abstract}

\begin{IEEEkeywords}
Multiple target filtering, point targets, extended targets. 
\end{IEEEkeywords}

\section{Introduction}

\label{sec:Introduction}

Multiple target filtering refers to the sequential estimation of the
states of the current targets, which may appear, move and disappear,
given past and current noisy sensor measurements. This is a key component
in many applications such as self-driving vehicles \cite{Choi13}
and maritime navigation \cite{Brekke19}. Multi-target filtering can
be solved in a Bayesian framework by computing the posterior density
on the current set of targets, given probabilistic models for target
births, dynamics and deaths, and also models for the measurements,
obtained from one or multiple sensors \cite{Mahler_book14,Saucan17}.
The target birth model contains probabilistic information on where
targets may appear in the surveillance area, and it enables the resulting
filters to contain information on potential targets that may remain
occluded \cite[Fig. 6]{Granstrom20}, which is of paramount importance
in some applications such as self-driving vehicles.

If the target extent is small compared to the sensor resolution, it
is common to use point-target modelling. In this model, a target state
typically contains kinematic information, such as position and velocity,
and one target can generate at most one measurement at each time step
\cite{Blackman04}. Conversely, if the target extent is large compared
to the sensor resolution, a better choice is to use extended target
modelling \cite{Granstrom17}. Here, the target state usually contains
both kinematic information and information on its extent, e.g., represented
by an ellipse \cite{Koch08}. In addition, each extended target may
generate more than one measurement at each time step, represented
via a Poisson point process (PPP) in the standard model \cite{Gilholm05,Granstrom17,Tang19}.

For both point and extended targets with Poisson birth model and the
standard measurement models, the posterior density is a Poisson multi-Bernoulli
mixture (PMBM), which can be calculated by the corresponding PMBM
filtering recursions\footnote{A course on multiple target tracking with detailed information on
these topics can be found at https://www.youtube.com/channel/UCa2-fpj6AV8T6JK1uTRuFpw.} \cite{Williams15b,Angel18_b,Granstrom20}. The PMBM has a compact
representation of global hypotheses, representing undetected targets
via the intensity of a PPP and making use of probabilistic target
existence in each global hypothesis. The PMBM recursion can also handle
a multi-Bernoulli birth model by setting the PPP intensity to zero
and adding new Bernoulli components in the prediction step, resulting
in the MBM filter \cite{Angel18_b,Angel19_e}. The MBM filter can
also be extended to consider multi-Bernoullis with deterministic target
existence, which we refer to as the MBM$_{01}$ filter, at the expense
of increasing the number of global hypotheses \cite[Sec. IV]{Angel18_b}.
Both MBM and MBM$_{01}$ filters can consider target states with labels,
and the (labelled) MBM$_{01}$ filtering recursion is analogous to
the $\delta$-generalised labelled multi-Bernoulli ($\delta$-GLMB)
filter \cite{Vo13,Beard16}. 

There are applications in which it is important to have more general
models than the standard point and extended target models \cite{Mahler_book14}.
Specifically, there may be some targets that are small compared to
the sensor resolution, while other targets are large, which implies
that there are coexisting point/extended targets in the field of view.
For example, in a self-driving vehicle application, pedestrians may
be modelled as point targets while other vehicles as extended targets.
The distinction between point and extended targets may also depend
on the distance, as sensor resolution is usually higher at short distances.
Therefore, it is of interest to develop multi-target filters that
can handle coexisting point and extended targets. The extended target
measurement models in \cite{Lundquist13,Beard16} are general enough
to model measurements from coexisting point and extended targets,
but no single-target state, dynamic model and filter implementations
are presented for this case.

In this paper, we fill this gap and propose a PMBM filter for coexisting
point and extended targets. In order to do so, we first develop a
PMBM filtering recursion for a generalised measurement model, in which
each target generates an independent set of measurements with an arbitrary
distribution, and clutter is a PPP. With a suitable choice of the
target-generated measurement distribution, this generalised model
recovers the standard point and extended target measurement models.
As a result, the PMBM filter with the generalised measurement model
can be used to address multi-target filtering problems with point
and extended targets \cite{Williams15b,Angel18_b,Granstrom20}, and
more general problems. For example, the generalised measurement model
can also be used for diffuse multipath \cite{Ge20}, extended targets
composed of point-scatterers \cite{Granstrom17}, and point targets
with stationary landmarks, modelled as extended targets. The resulting
PMBM recursion has a track-oriented form that enables efficient implementation
\cite{Kurien_inbook90}.

Based on the developed PMBM filtering recursion, the second contribution
is to derive a PMBM filter for coexisting point and extended targets.
In this setting, each Bernoulli contains probabilistic information
on target existence and type, either point or extended target. The
implementation is provided for a linear Gaussian model for point targets
\cite{Sarkka_book13} and a Gamma Gaussian Inverse Wishart (GGIW)
model for extended targets \cite{Granstrom15,Granstrom17,Granstrom20}.
Finally, we explain how a PMBM density in this context can be projected
onto a Poisson multi-Bernoulli (PMB) density \cite{Williams15b}.
Performing this projection after each update provides us with a PMB
filter, which is a fast approximation to the PMBM filter. Simulation
results are provided to analyse the performance of the filters.

The rest of the paper is organised as follows. Section \ref{sec:Problem_formulation}
introduces the problem formulation and an overview of the solution.
The update for the PMBM filter with generalised measurement model
is derived in Section \ref{sec:PMBM-update}. Section \ref{sec:Coexisting_point_extended}
explains the PMBM filter for coexisting point and extended targets,
and the PMB projection. Simulation results and conclusions are given
in Sections \ref{sec:Simulations} and \ref{sec:Conclusions}, respectively.

\section{Problem formulation and overview of the solution}

\label{sec:Problem_formulation}

This paper deals with multiple target tracking with both point and
extended target models. This section presents an overview of a PMBM
filter with a generalised measurement model that will be used to model
coexisting point and extended targets in Section \ref{sec:Coexisting_point_extended}.
We introduce the models in Section \ref{subsec:Models} and the PMBM
filter overview in Section \ref{subsec:PMBM-filter-overview}.

\subsection{Models\label{subsec:Models}}

A single target state $x\in\mathcal{X}$, where $\mathcal{X}$ is
a locally compact, Hausdorff and second-countable (LCHS) space \cite{Mahler_book14},
contains the information of interest about the target, for example,
its position, velocity and extent. The set of targets at time $k$
is denoted by $X_{k}\in\mathcal{F}\left(\mathcal{X}\right)$, where
$\mathcal{F}\left(\mathcal{X}\right)$ represents the set of finite
subsets of $\mathcal{X}$. 

A main novelty in this paper is the development of a PMBM filter with
a generalised measurement model. Here, the set $X_{k}$ of targets
at time step $k$, is observed through a set $Z_{k}\in\mathcal{F}\left(\mathbb{R}^{n_{z}}\right)$
of noisy measurements, which consist of the union of target-generated
measurements and clutter, with the model:
\begin{itemize}
\item Each target $x\in X_{k}$ generates an independent set $Z$ of measurements
with density $f\left(Z|x\right)$. 
\item Clutter is a PPP with intensity $\lambda^{C}\left(\cdot\right)$. 
\end{itemize}
It should be noted that the standard point and extended measurement
models \cite{Williams15b,Granstrom20} can be recovered by suitable
choices of $f\left(Z|x\right)$. 

We also consider the standard dynamic model for targets. Given the
set $X_{k}$ of targets at time step $k$, each target $x\in X_{k}$
survives with probability $p^{S}\left(x\right)$ and moves to a new
state with a transition density $g\left(\cdot\left|x\right.\right)$,
or dies with probability $1-p^{S}\left(x\right)$. At time step $k$,
targets are born independently following a Poisson point process (PPP)
with intensity $\lambda_{k}^{B}\left(\cdot\right)$. 

\subsection{PMBM posterior\label{subsec:PMBM-filter-overview}}

In this paper, we show that, for the above-mentioned measurement and
dynamic models, the density $f_{k|k'}\left(\cdot\right)$ of $X_{k}$
given the sequence of measurements $\left(Z_{1},...,Z_{k'}\right)$,
where $k'\in\left\{ k-1,k\right\} $, is a PMBM density. This section
provides an overview of the PMBM posterior and its data association
hypotheses. 

The PMBM is of the form \cite{Williams15b,Angel18_b}
\begin{align}
f_{k|k'}\left(X_{k}\right) & =\sum_{Y\uplus W=X_{k}}f_{k|k'}^{\mathrm{p}}\left(Y\right)f_{k|k'}^{\mathrm{mbm}}\left(W\right)\label{eq:PMBM}\\
f_{k|k'}^{\mathrm{p}}\left(X_{k}\right) & =e^{-\int\lambda_{k|k'}\left(x\right)dx}\prod_{x\in X_{k}}\lambda_{k|k'}\left(x\right)\\
f_{k|k'}^{\mathrm{mbm}}\left(X_{k}\right) & =\sum_{a\in\mathcal{A}_{k|k'}}w_{k|k'}^{a}\sum_{\uplus_{l=1}^{n_{k|k'}}X^{l}=X_{k}}\prod_{i=1}^{n_{k|k'}}f_{k|k'}^{i,a^{i}}\left(X^{i}\right)
\end{align}
where $\lambda_{k|k'}\left(\cdot\right)$ is the intensity of the
PPP $f_{k|k'}^{\mathrm{p}}\left(\cdot\right)$, representing undetected
targets, and $f_{k|k'}^{\mathrm{mbm}}\left(\cdot\right)$ is a multi-Bernoulli
mixture representing potential targets that have been detected at
some point up to time step $k'$. Symbol $\uplus$ denotes the disjoint
union and the summation in \eqref{eq:PMBM} is taken over all mutually
disjoint (and possibly empty) sets $Y$ and $W$ whose union is $X_{k}$,
i.e., $X_{k}$ is fixed, and $Y$ and $W$ free. 

In the PMBM posterior, there are $n_{k|k'}$ Bernoulli components
and for each Bernoulli there are $h_{k|k'}^{i}$ possible local hypotheses.
By selecting a local hypothesis $a^{i}\in\left\{ 1,...,h_{k|k'}^{i}\right\} $
for each Bernoulli, we obtain a global hypothesis $a=\left(a^{1},...,a^{n_{k|k'}}\right)\in\mathcal{A}_{k|k'}$,
where $\mathcal{A}_{k|k'}$ is the set of global hypotheses. Each
global hypothesis represents a multi-Bernoulli distribution. The $i$-th
Bernoulli component with local hypothesis $a^{i}$ has a density
\begin{align}
f_{k|k'}^{i,a^{i}}\left(X\right) & =\begin{cases}
1-r_{k|k'}^{i,a^{i}} & X=\emptyset\\
r_{k|k'}^{i,a^{i}}f_{k|k'}^{i,a^{i}}\left(x\right) & X=\left\{ x\right\} \\
0 & \mathrm{otherwise}
\end{cases}\label{eq:Bernoulli_density_filter}
\end{align}
where $r_{k|k'}^{i,a^{i}}$ is the probability of existence and $f_{k|k'}^{i,a^{i}}\left(x\right)$
the single target density. The weight of global hypothesis $a$ is
$w_{k|k'}^{a}$ and meets
\begin{equation}
w_{k|k'}^{a}\propto\prod_{i=1}^{n_{k|k'}}w_{k|k'}^{i,a^{i}}\label{eq:global_weights}
\end{equation}
where $w_{k|k'}^{i,a^{i}}$ is the weight of the $i$-th Bernoulli
with local hypothesis $a^{i}$, and $\sum_{a\in\mathcal{A}_{k|k'}}w_{k|k'}^{a}=1$. 

The set of feasible global hypotheses is defined as in the extended
target case \cite{Granstrom20,Xia19}. We denote the measurement set
at time step $k$ as $Z_{k}=\left\{ z_{k}^{1},...,z_{k}^{m_{k}}\right\} $.
We refer to measurement $z_{k}^{j}$ using the pair $\left(k,j\right)$
and the set of all such measurement pairs up to (and including) time
step $k$ is denoted by $\mathcal{M}_{k}$. Then, a single target
hypothesis $a^{i}$ for the $i$-th Bernoulli component has a set
of measurement pairs denoted as $\mathcal{M}_{k}^{i,a^{i}}\subseteq\mathcal{M}_{k}$.
The set $\mathcal{A}_{k|k'}$ of all global hypotheses meets
\begin{align*}
\mathcal{A}_{k|k'}= & \left\{ \left(a^{1},...,a^{n_{k|k'}}\right):a^{i}\in\left\{ 1,...,h_{k|k'}^{i}\right\} \,\forall i,\right.\\
 & \left.\bigcup_{i=1}^{n_{k|k'}}\mathcal{M}_{k'}^{i,a^{i}}=\mathcal{M}_{k'},\mathcal{M}_{k'}^{i,a^{i}}\cap\mathcal{M}_{k'}^{j,a^{j}}=\emptyset,\,\forall i\neq j\right\} .
\end{align*}
That is, all measurements must be assigned to a local hypothesis,
and there cannot be more than one local hypothesis with the same measurement.
More than one measurement can be associated to the same local hypothesis
at the same time step. Each global hypothesis therefore corresponds
to a unique partition of $\mathcal{M}_{k'}$ \cite[Sec. V]{Granstrom20},
and the number of global hypothesis is the Bell number of $\left|\mathcal{M}_{k'}\right|$.
At each time step, each non-empty subset of $Z_{k}$ generates a new
Bernoulli component, corresponding to a potential target detected
for the first time or clutter. This implies that, at each time step,
$2^{m_{k}}-1$ new Bernoulli components are generated.  

It should be noted that the prediction step of a PMBM density is closed-form
for the standard dynamic models \cite{Williams15b,Angel18_b}, and
is not affected by the choice of measurement model. Therefore, the
next section focuses on the update  and we omit the details for prediction,
which can be found in \cite{Williams15b,Angel18_b}.

\section{PMBM update for a generalised measurement model\label{sec:PMBM-update}}

This section provides the PMBM filter update step with the measurement
model in Section \ref{sec:Problem_formulation}. We denote a Kronecker
delta as $\delta_{i}\left[\cdot\right]$, with $\delta_{i}\left[u\right]=1$
if $u=i$ and $\delta_{i}\left[u\right]=0$, otherwise. Also, given
two real-valued functions $a\left(\cdot\right)$ and $b\left(\cdot\right)$
on the target space, we denote their inner product as
\begin{align}
\left\langle a,b\right\rangle  & =\int a\left(x\right)b\left(x\right)dx.
\end{align}

\subsection{Update}

The update of the predicted PMBM $f_{k|k-1}\left(\cdot\right)$ after
observing $Z_{k}$ is given in the following theorem.
\begin{thm}
\label{thm:PMBM_update}Assume the predicted density $f_{k|k-1}\left(\cdot\right)$
is a PMBM of the form \eqref{eq:PMBM}. Then, the updated density
$f_{k|k}\left(\cdot\right)$ with set $Z_{k}=\left\{ z_{k}^{1},...,z_{k}^{m_{k}}\right\} $
is a PMBM with the following parameters. The number of Bernoulli components
is $n_{k|k}=n_{k|k-1}+2^{m_{k}}$. The intensity of the PPP is
\begin{align}
\lambda_{k|k}\left(x\right) & =f\left(\emptyset|x\right)\lambda_{k|k-1}\left(x\right).\label{eq:updated_PPP}
\end{align}
For Bernoullis continuing from previous time steps $i\in\left\{ 1,...,n_{k|k-1}\right\} $,
a new local hypothesis is included for each previous local hypothesis
and either a misdetection or an update with a non-empty subset of
$Z_{k}$. The updated number of local hypotheses is $h_{k|k}^{i}=2^{m_{k}}h_{k|k-1}^{i}$.
For missed detection hypotheses, $i\in\left\{ 1,...,n_{k|k-1}\right\} $,
$a^{i}\in\left\{ 1,...,h_{k|k-1}^{i}\right\} $, we obtain
\begin{align}
\mathcal{M}_{k}^{i,a^{i}} & =\mathcal{M}_{k-1}^{i,a^{i}}\label{eq:Miss_measurement}\\
l_{k|k}^{i,a^{i},\emptyset} & =\big\langle f_{k|k-1}^{i,a^{i}},f\left(\emptyset|\cdot\right)\big\rangle\label{eq:Miss_likelihood}\\
w_{k|k}^{i,a^{i}} & =w_{k|k-1}^{i,a^{i}}\left[1-r_{k|k-1}^{i,a^{i}}+r_{k|k-1}^{i,a^{i}}l_{k|k}^{i,a^{i},\emptyset}\right]\label{eq:Miss_weight}\\
r_{k|k}^{i,a^{i}} & =\frac{r_{k|k-1}^{i,a^{i}}l_{k|k}^{i,a^{i},\emptyset}}{1-r_{k|k-1}^{i,a^{i}}+r_{k|k-1}^{i,a^{i}}l_{k|k}^{i,a^{i},\emptyset}}\label{eq:Miss_existence}\\
f_{k|k}^{i,a^{i}}(x) & =\frac{f\left(\emptyset|x\right)f_{k|k-1}^{i,a^{i}}(x)}{l_{k|k}^{i,a^{i},\emptyset}}.\label{eq:Miss_density}
\end{align}

Let $Z_{k}^{1},...,Z_{k}^{2^{m_{k}}-1}$ be the nonempty subsets of
$Z_{k}$. For a Bernoulli $i\in\left\{ 1,...,n_{k|k-1}\right\} $
with a single target hypothesis $\widetilde{a}^{i}\in\left\{ 1,...,h_{k|k-1}^{i}\right\} $
in the predicted density, the new local hypothesis generated by a
set $Z_{k}^{j}$ has $a^{i}=\widetilde{a}^{i}+h_{k|k-1}^{i}j$, $r_{k|k}^{i,a^{i}}=1$,
and
\begin{align}
\mathcal{M}_{k}^{i,a^{i}} & =\mathcal{M}_{k-1}^{i,\widetilde{a}^{i}}\cup\left\{ \left(k,p\right):z_{k}^{p}\in Z_{k}^{j}\right\} \\
l_{k|k}^{i,a^{i},Z_{k}^{j}} & =\bigg\langle f_{k|k-1}^{i,\widetilde{a}^{i}},f\left(Z_{k}^{j}|\cdot\right)\bigg\rangle\\
w_{k|k}^{i,a^{i}} & =w_{k|k-1}^{i,\widetilde{a}^{i}}r_{k|k-1}^{i,\widetilde{a}^{i}}l_{k|k}^{i,a^{i},Z_{k}^{j}}\label{eq:update_weight}\\
f_{k|k}^{i,a^{i}}(x) & =\frac{f\left(Z_{k}^{j}|x\right)f_{k|k-1}^{i,\widetilde{a}^{i}}(x)}{l_{k|k}^{i,a^{i},Z_{k}^{j}}}.\label{eq:update_density}
\end{align}
For the new Bernoulli initiated by subset $Z_{k}^{j}$, whose index
is $i=n_{k|k-1}+j$, we have two single target hypotheses ($h_{k|k}^{i}=2$),
one corresponding to a non-existent Bernoulli
\begin{equation}
\mathcal{M}_{k}^{i,1}=\emptyset,\;w_{k|k}^{i,1}=1,\;r_{k|k}^{i,1}=0
\end{equation}
and the other
\begin{align}
\mathcal{M}_{k}^{i,2} & =\left\{ \left(k,p\right):z_{k}^{p}\in Z_{k}^{j}\right\} \\
l_{k|k}^{Z_{k}^{j}} & =\bigg\langle\lambda_{k|k-1},f\left(Z_{k}^{j}|\cdot\right)\bigg\rangle\\
w_{k|k}^{i,2} & =\delta_{1}\left[|Z_{k}^{j}|\right]\left[\prod_{z\in Z_{k}^{j}}\lambda^{C}\left(z\right)\right]+l_{k|k}^{Z_{k}^{j}}\label{eq:new_Bernoulli_weight}\\
r_{k|k}^{i,2} & =\frac{l_{k|k}^{Z_{k}^{j}}}{w_{k|k}^{i,a^{i}}}\label{eq:new_Bernoulli_existence}\\
f_{k|k}^{i,2}(x) & =\frac{f\left(Z_{k}^{j}|x\right)\lambda_{k|k-1}(x)}{l_{k|k}^{Z_{k}^{j}}}.\quad\square\label{eq:new_Bernoulli_density}
\end{align}
\end{thm}
Theorem \ref{thm:PMBM_update} is proved in Appendix \ref{sec:Update_proof_App}.
We can see that the updated PPP intensity in \eqref{eq:updated_PPP}
corresponds to the predicted intensity multiplied by the probability
of not receiving any measurements. This is expected as the PPP contains
information on the undetected targets. Misdetection hypotheses lower
the probability of existence of the Bernoullis via \eqref{eq:Miss_likelihood}
and \eqref{eq:Miss_existence}. If $f\left(\emptyset|x\right)$ does
not depend on $x$, the single-target densities of misdetection hypotheses
remain unchanged, see \eqref{eq:Miss_density}.

For the update of a previous Bernoulli component with subset $Z_{k}^{j}$,
the updated Bernoulli has a probability of existence equal to one.
Each non-empty subset $Z_{k}^{j}\subseteq Z_{k}$ creates a new Bernoulli
component. If $|Z_{k}^{j}|>1$, the existence probability $r_{k|k}^{i,2}$
of the new Bernoulli component is one, which implies that, conditioned
on the corresponding hypothesis, this Bernoulli represents an existing
target. If $|Z_{k}^{j}|=1$, the existence probability $r_{k|k}^{i,2}$
of the new Bernoulli component depends on the clutter intensity $\lambda^{C}\left(\cdot\right)$,
as this Bernoulli may correspond to a target or to clutter. The higher
$\lambda^{C}\left(\cdot\right)$, the lower the probability of existence
of this potential target. 

\subsection{Relation to standard point/extended target models\label{subsec:Relation-to-standard-point-extended-models}}

In the standard point target measurement model, a target $x$ is detected
with probability $p^{D}\left(x\right)$ and, if detected, it generates
one measurement with density $l(\cdot|x)$. This model is obtained
by setting
\begin{align}
f\left(Z|x\right) & =\begin{cases}
1-p^{D}\left(x\right) & Z=\emptyset\\
p^{D}\left(x\right)l(z|x) & Z=\left\{ z\right\} \\
0 & \left|Z\right|>1.
\end{cases}\label{eq:standard_point_measurement}
\end{align}
If we use the above definitions of local and global hypotheses and
Theorem \ref{thm:PMBM_update} for point targets, many of the global
hypotheses contain local hypotheses where more than one measurement
is associated to the same Bernoulli at the same time step. Since this
is impossible according to \eqref{eq:standard_point_measurement},
all these hypotheses would obtain weight zero. A more convenient way
to handle point targets is to exclude these hypotheses from the set
$\mathcal{A}_{k|k}$ that we consider, see \cite{Williams15b}. 

In the standard extended target model, a target $x$ is detected with
probability $p^{D}\left(x\right)$ and, if detected, it generates
a PPP measurement with intensity $\gamma\left(x\right)l(\cdot|x)$,
where $l(\cdot|x)$ is a single-measurement density and $\gamma\left(x\right)$
is the expected number of measurements. We can recover this model
by setting

\begin{align}
f\left(Z|x\right) & =\begin{cases}
1-p^{D}\left(x\right)+p^{D}\left(x\right)e^{-\gamma\left(x\right)} & Z=\emptyset\\
p^{D}\left(x\right)\gamma^{\left|Z\right|}\left(x\right)e^{-\gamma\left(x\right)}\prod_{z\in Z}l(z|x) & \left|Z\right|>0.
\end{cases}\label{eq:standard_extended_measurement}
\end{align}
In this case, Theorem \ref{thm:PMBM_update} becomes the standard
extended-target PMBM update in track-oriented form \cite{Granstrom20,Xia19}.

\subsection{Discussion\label{subsec:Discussion}}

We have shown that the update of a PMBM prior with the generalised
measurement model in Section \ref{sec:Problem_formulation} is also
PMBM. The proposed measurement model contains the standard point target
and extended target measurement models as particular cases, and can
be used for other types of measurement modelling. For example, another
important special case is that each target could generate a union
of independent Bernoulli measurements, which can model extended targets
that consist of reflection points \cite{Broida90,Hammarstrand12}.
It can also model extended targets with binomially distributed target-generated
measurements \cite{Ristic13b}. The considered measurement model also
allows us to model coexisting point and extended targets, for example,
modelling radar returns from vehicles (extended targets) and pedestrians
(point targets), as will be explained in Section \ref{sec:Coexisting_point_extended}.
It can also model scenarios in which far-away targets produce point-target
measurements and targets that are sufficiently close produce extended-target
measurements, for example, by setting a distance threshold, which
may depend on the target extent, to switch between both types of model.
The proposed PMBM update requires PPP clutter, which can be relaxed
in Bernoulli filters \cite{Shen18}.

We would also like to remark that we have presented the results for
PPP birth density, as we think this is generally the most suitable
birth process, due to the lower number of generated hypotheses \cite{Angel18_b,Angel19_e}.
Nevertheless, the presented results also hold for the following cases.
For multi-Bernoulli birth, the above equations are valid, by setting
the Poisson intensity equal to zero, and adding the Bernoulli components
for new born targets in the prediction step \cite{Angel18_b,Angel19_e}.
In this case, the posterior is a multi-Bernoulli mixture (MBM), which
can also be represented as $\mathrm{MBM}_{01}$ \cite[Sec. IV]{Angel18_b}.
For multi-Bernoulli birth, one can also uniquely label each Bernoulli
component, for which the labelled $\mathrm{MBM}_{01}$ recursion would
correspond to the $\delta$-GLMB filter recursion \cite{Beard16}.

\section{PMBM filter for coexisting point and extended targets\label{sec:Coexisting_point_extended}}

This section presents the PMBM filter, and a track-oriented PMB filter,
for coexisting point and extended targets.  The single target space
for point targets is $\mathbb{R}^{n_{x}}$, which represents the kinematic
state (e.g. position and velocity). We model extended targets with
the GGIW model \cite{Lundquist13}, whose space is $\mathcal{X}_{e}=\mathbb{R}_{+}\times\mathbb{R}^{n_{x}}\times\mathbb{S}_{+}^{d}$,
where $\mathbb{R}_{+}$ represents the positive real numbers and $\mathbb{S}_{+}^{d}$
the positive definite matrices of size $d$, which is the dimension
of the extent. 

The single target space for coexisting point/extended targets is then
$\mathcal{X}=\mathbb{R}^{n_{x}}\uplus\mathcal{X}_{e}$, where $\uplus$
stands for union of sets that are mutually disjoint, i.e., $\mathcal{X}=\mathbb{R}^{n_{x}}\cup\mathcal{X}_{e}$
and $\mathbb{R}^{n_{x}}\cap\mathcal{X}_{e}=\emptyset$ \cite{Mahler_book14}.
Other works with this type of hybrid space are for example \cite{Mahler_book14,Mahler11,Angel20_b,Xia19_b}.
If $x\in\mathcal{X}_{e}$, then $x=\left(\gamma,\xi,X\right)$, where
$\gamma$ represents the expected number of measurements per target,
$\xi$ is the kinematic state and $X$ is the extent state that describes
the target's size and shape. It should be noted that, though not necessary,
it is also possible to include a class variable in the target space
to distinguish between point and extended targets, as in interacting
multiple models \cite{Mazor98}, see Appendix \ref{sec:IMM_relation_appendix}.
This appendix also explains the corresponding single-target integral.

We use a measurement model that corresponds to the standard point
and extended target measurement models depending on the type of target
we observe. That is, for $x\in\mathbb{R}^{n_{x}}$, $f(Z|x)$ is given
by \eqref{eq:standard_point_measurement} with a probability $p^{D}\left(x\right)=p_{1}^{D}$
of detection, $l(z|x)=\mathcal{N}\left(z;H_{1}x,R\right)$ where $H_{1}$
is the measurement matrix, $R$ is the noise covariance matrix, and
$\mathcal{N}\left(\cdot;\overline{x},P\right)$ is a Gaussian density
with mean $\overline{x}$ and covariance $P$. For $x\in\mathcal{X}_{e}$,
$f(Z|x)$ is given by \eqref{eq:standard_extended_measurement} with
a probability $p^{D}\left(x\right)=p_{2}^{D}$ of detection, $\gamma\left(x\right)=\gamma$,
and $l(z|x)=\mathcal{N}\left(z;H_{2}\xi,X\right)$ where $H_{2}$
is the measurement matrix.

The rest of this section is organised as follows. Section \ref{subsec:Single-target-densities}
presents the considered single-target densities. The update and the
prediction are provided in Sections \ref{subsec:Update_coexisting}
and \ref{subsec:Prediction_coexisting}. The PMB approximation is
addressed in Section \ref{subsec:PMB-approximation}. Target state
estimation is explained in Section \ref{subsec:Estimation}. Practical
aspects are discussed in Section \ref{subsec:Practical-aspects}.

\subsection{Single-target densities\label{subsec:Single-target-densities}}

We develop a PMBM implementation in which we propagate a Gaussian
for single target densities and a (factorised) GGIW density for extended
target densities \cite{Granstrom15,Granstrom17,Feldmann11}. In a
factorised GGIW density, the distributions for $\gamma$, $\xi$ and
$X$ are independent, which has computational and practical benefits
\cite[Sec. III.A.2]{Granstrom17}. 

The Gaussian density for $x\in\mathcal{X}$ with mean $\overline{x}_{k|k'}^{i,a^{i},1}$
and covariance matrix $P_{k|k'}^{i,a^{i},1}$ is
\begin{align}
\mathcal{N}_{p}\left(x;\overline{x}_{k|k'}^{i,a^{i},1},P_{k|k'}^{i,a^{i},1}\right) & =\mathcal{N}\left(x;\overline{x}_{k|k'}^{i,a^{i},1},P_{k|k'}^{i,a^{i},1}\right)
\end{align}
for $x\in\mathbb{R}^{n_{x}}$ and zero for $x\in\mathcal{X}_{e}$.
 Note that $\mathcal{N}_{p}\left(\cdot\right)$ is zero evaluated
at $x\in\mathcal{X}_{e}$, as $\mathcal{N}_{p}\left(\cdot\right)$
represents point targets. 

The Gamma density with parameters $\alpha>0$ and $\beta>0$ is denoted
as $\mathcal{G}\left(\cdot;\alpha,\beta\right)$. The inverse Wishart
density on matrices in $\mathbb{S}_{+}^{d}$ with $v>2d$ degrees
of freedom and parameter matrix $V\in\mathbb{S}_{+}^{d}$ is denoted
as $\mathcal{IW}\left(;v,V\right)$ \cite{Gupta_book99}. Then, the
GGIW density for $x\in\mathcal{X}$ with parameters
\begin{align}
\zeta_{k|k'}^{i,a^{i}} & =\left(\alpha_{k|k'}^{i,a^{i}},\beta_{k|k'}^{i,a^{i}},\overline{x}_{k|k'}^{i,a^{i},2},P_{k|k'}^{i,a^{i},2},v_{k|k'}^{i,a^{i}},V_{k|k'}^{i,a^{i}}\right)
\end{align}
is
\begin{align}
\mathcal{G}_{e}\left(x;\zeta_{k|k'}^{i,a^{i}}\right) & =\mathcal{G}\left(\gamma;\alpha_{k|k'}^{i,a^{i}},\beta_{k|k'}^{i,a^{i}}\right)\mathcal{N}\left(\xi;\overline{x}_{k|k'}^{i,a^{i},2},P_{k|k'}^{i,a^{i},2}\right)\nonumber \\
 & \times\mathcal{IW}\left(X;v_{k|k'}^{i,a^{i}},V_{k|k'}^{i,a^{i}}\right)\label{eq:GGIW}
\end{align}
for $x\in\mathcal{X}_{e}$ and zero for $x\in\mathbb{R}^{n_{x}}$.

The single-target density of the $i$-th Bernoulli and local hypothesis
$a^{i}$ is
\begin{align}
f_{k|k'}^{i,a^{i}}\left(x\right) & =c_{k|k'}^{i,a^{i}}\mathcal{N}_{p}\left(x;\overline{x}_{k|k'}^{i,a^{i},1},P_{k|k'}^{i,a^{i},1}\right)\nonumber \\
 & \quad+\left(1-c_{k|k'}^{i,a^{i}}\right)\mathcal{G}_{e}\left(x;\zeta_{k|k'}^{i,a^{i}}\right)\label{eq:single_target_joint_Gaussian}
\end{align}
where $c_{k|k'}^{i,a^{i}}$ and $\left(1-c_{k|k'}^{i,a^{i}}\right)$
are the probabilities that the target is a point-target and  extended
target, respectively. The PPP intensity is a mixture
\begin{align}
\lambda_{k|k'}\left(x\right) & =\sum_{q=1}^{n_{k|k'}^{p}}w_{k|k'}^{p,q}\mathcal{\mathcal{N}}_{p}\left(x;\overline{x}_{k|k'}^{p,q,1},P_{k|k'}^{p,q,1}\right)\nonumber \\
 & \quad+\sum_{q=1}^{n_{k|k'}^{e}}w_{k|k'}^{e,q}\mathcal{G}_{e}\left(x;\zeta_{k|k'}^{e,q}\right)\label{eq:PPP_joint_Gaussian}
\end{align}
where $n_{k|k'}^{p}$ is the number of components with point-targets,
with weight $w_{k|k'}^{p,q}$, mean $\overline{x}_{k|k'}^{p,q,1}$
and covariance $P_{k|k'}^{p,q,1}$, and $n_{k|k'}^{e}$ is the number
of components with extended targets, with weight $w_{k|k'}^{e,q}$
and parameters $\zeta_{k|k'}^{e,q}$. It should be noted that $\sum_{q=1}^{n_{k|k'}^{p}}w_{k|k'}^{p,q}$
and $\sum_{q=1}^{n_{k|k'}^{e}}w_{k|k'}^{e,q}$ represent the expected
number of undetected point and extended targets, respectively.

\subsection{Update\label{subsec:Update_coexisting}}

We represent the update of a GGIW density with parameters $\zeta_{k|k-1}^{i,a^{i}}$
with a given measurement set $Z_{k}^{j}$ as a function
\begin{align*}
\left(\zeta_{k|k}^{e,q},\ell_{k|k}^{e,q}\right) & =\mathrm{u}_{e}\left(\zeta_{k|k-1}^{i,a^{i}},Z_{k}^{j}\right)
\end{align*}
where $\zeta_{k|k}^{e,q}$ is the updated GGIW and $\ell_{k|k}^{e,q}$
the marginal likelihood, see Appendix \ref{sec:Update_rules_appendix}.
The Kalman filter update of a Gaussian density with mean $\overline{x}_{k|k-1}^{i,a^{i},1}$
and covariance $P_{k|k-1}^{i,a^{i},1}$ and measurement $z$ is represented
as
\begin{align*}
\left(\overline{x}_{k|k}^{i,a^{i},1},P_{k|k}^{i,a^{i},1},\ell_{k|k}^{i,a^{i},1}\right) & =\mathrm{u}_{p}\left(\overline{x}_{k|k-1}^{i,a^{i},1},P_{k|k-1}^{i,a^{i},1},z\right)
\end{align*}
where $\overline{x}_{k|k}^{i,a^{i},1}$ and $P_{k|k}^{i,a^{i},1}$
are the updated mean and covariance, and $\ell_{k|k}^{i,a^{i},1}$
is the marginal likelihood, see \cite{Sarkka_book13} for details. 

We apply Theorem \ref{thm:PMBM_update} to obtain the specific parameters
of the updated PMBM provided in the following lemma. 
\begin{lem}
\label{lem:PMBM_update_Gaussian} The updated PMBM with a prior PMBM
described by \eqref{eq:PMBM}, \eqref{eq:single_target_joint_Gaussian}
and \eqref{eq:PPP_joint_Gaussian}, with measurement set $Z_{k}=\left\{ z_{k}^{1},...,z_{k}^{m_{k}}\right\} $
has the structure in Theorem \ref{thm:PMBM_update} with the following
parameters. The number of PPP components is $n_{k|k}^{p}=n_{k|k-1}^{p}$
and $n_{k|k}^{e}=2n_{k|k-1}^{e}$. For point targets,
\begin{equation}
\overline{x}_{k|k}^{p,q,1}=\overline{x}_{k|k-1}^{p,q,1},P_{k|k}^{p,q,1}=P_{k|k-1}^{p,q,1},
\end{equation}
\begin{equation}
w_{k|k}^{p,q}=\left(1-p_{1}^{D}\right)w_{k|k-1}^{p,q}.
\end{equation}

For extended targets and $q\leq n_{k|k-1}^{e}$, we have
\begin{equation}
\zeta_{k|k}^{e,q}=\zeta_{k|k-1}^{e,q},w_{k|k}^{e,q}=\left(1-p_{2}^{D}\right)w_{k|k-1}^{e,q}.
\end{equation}
For $q>n_{k|k-1}^{e}$, $\tilde{q}=q-n_{k|k-1}^{e}$,
\begin{align}
\left(\zeta_{k|k}^{e,q},\ell_{k|k}^{e,q}\right) & =\mathrm{u}_{e}\left(\zeta_{k|k-1}^{e,\tilde{q}},\emptyset\right)\\
w_{k|k}^{e,q} & =p_{2}^{D}\ell_{k|k}^{e,q}w_{k|k}^{e,\tilde{q}}.
\end{align}

For missed detection hypotheses of previous Bernoullis,
\begin{align}
f_{k|k}^{i,a^{i}}\left(x\right) & =c_{k|k}^{i,a^{i}}\mathcal{N}_{p}\left(x;\overline{x}_{k|k}^{i,a^{i},1},P_{k|k}^{i,a^{i},1}\right)+\left(1-c_{k|k}^{i,a^{i}}\right)\nonumber \\
 & \quad\times\left[w\mathcal{G}_{e}\left(x;\zeta_{k|k}^{i,a^{i},1}\right)+\left(1-w\right)\mathcal{G}_{e}\left(x;\zeta_{k|k}^{i,a^{i},2}\right)\right]\label{eq:missed_Bernoulli_update_Gaussian}
\end{align}
where $\overline{x}_{k|k}^{i,a^{i},1}=\overline{x}_{k|k-1}^{i,a^{i},1}$,
$P_{k|k}^{i,a^{i},1}=P_{k|k-1}^{i,a^{i},1}$, $\zeta_{k|k}^{i,a^{i},1}=\zeta_{k|k-1}^{i,a^{i}}$
and
\begin{align}
\left(\zeta_{k|k}^{i,a^{i},2},\ell_{k|k}^{i,a^{i},2}\right) & =\mathrm{u}_{e}\left(\zeta_{k|k-1}^{i,a^{i}},\emptyset\right)\\
l_{k|k}^{i,a^{i},\emptyset} & =c_{k|k-1}^{i,a^{i}}\left(1-p_{1}^{D}\right)\nonumber \\
 & \,+\left(1-c_{k|k-1}^{i,a^{i}}\right)\left(1-p_{2}^{D}+p_{2}^{D}\ell_{k|k}^{i,a^{i},2}\right)\\
w_{k|k}^{i,a^{i}} & =w_{k|k-1}^{i,a^{i}}\left[1-r_{k|k-1}^{i,a^{i}}+r_{k|k-1}^{i,a^{i}}l_{k|k}^{i,a^{i},\emptyset}\right]\\
r_{k|k}^{i,a^{i}} & =\frac{r_{k|k-1}^{i,a^{i}}l_{k|k}^{i,a^{i},\emptyset}}{1-r_{k|k-1}^{i,a^{i}}+r_{k|k-1}^{i,a^{i}}l_{k|k}^{i,a^{i},\emptyset}}\\
c_{k|k}^{i,a^{i}} & =\frac{\left(1-p_{1}^{D}\right)c_{k|k-1}^{i,a^{i}}}{l_{k|k}^{i,a^{i},\emptyset}}\\
w & =\frac{1-p_{2}^{D}}{1-p_{2}^{D}+p_{2}^{D}\ell_{k|k}^{i,a^{i},2}}.
\end{align}
The detection hypotheses of a previous Bernoulli with a subset $Z_{k}^{j}$,
with $\left|Z_{k}^{j}\right|=m_{k}^{j}$, has $r_{k|k}^{i,a^{i}}=1$,
and
\begin{align}
w_{k|k}^{i,a^{i}} & =w_{k|k-1}^{i,\widetilde{a}^{i}}r_{k|k-1}^{i,\widetilde{a}^{i}}l_{k|k}^{i,a^{i},Z_{k}^{j}}\\
\left(\zeta_{k|k}^{i,a^{i}},\ell_{k|k}^{i,a^{i}}\right) & =\mathrm{u}_{e}\left(\zeta_{k|k-1}^{i,\widetilde{a}^{i}},Z_{k}^{j}\right).
\end{align}
For $m_{k}^{j}>1$, $l_{k|k}^{i,a^{i},Z_{k}^{j}}=p_{2}^{D}\ell_{k|k}^{i,a^{i}}$
and $c_{k|k}^{i,a^{i}}=0$, which implies that $\overline{x}_{k|k}^{i,a^{i},1}$
and $P_{k|k}^{i,a^{i},1}$ are irrelevant. For $m_{k}^{j}=1$, $Z_{k}^{j}=\left\{ z\right\} $,
we have
\begin{align}
\left(\overline{x}_{k|k}^{i,a^{i},1},P_{k|k}^{i,a^{i},1},\ell_{k|k}^{i,a^{i},1}\right) & =\mathrm{u}_{p}\left(\overline{x}_{k|k-1}^{i,a^{i},1},P_{k|k-1}^{i,a^{i},1},z\right)\\
l_{k|k}^{i,a^{i},Z_{k}^{j}} & =c_{k|k-1}^{i,a^{i}}p_{1}^{D}\ell_{k|k}^{i,a^{i},1}\nonumber \\
 & \quad+\left(1-c_{k|k-1}^{i,a^{i}}\right)p_{2}^{D}\ell_{k|k}^{i,a^{i}}\\
c_{k|k}^{i,a^{i}} & =\frac{c_{k|k-1}^{i,a^{i}}p_{1}^{D}\ell_{k|k}^{i,a^{i},1}}{l_{k|k}^{i,a^{i},Z_{k}^{j}}}.
\end{align}
For the new Bernoulli initiated by subset $Z_{k}^{j},$ the single
target density corresponding to an existing Bernoulli is
\begin{align}
f_{k|k}^{i,2}\left(x\right) & =c_{k|k}^{i,2}\sum_{q=1}^{n_{k|k-1}^{p}}w_{1}^{q}\mathcal{N}_{p}\left(x;\overline{x}_{k|k}^{i,2,q},P_{k|k}^{i,2,q}\right)\nonumber \\
 & \quad+\left(1-c_{k|k}^{i,2}\right)\sum_{q=1}^{n_{k|k-1}^{e}}w_{2}^{q}\mathcal{G}_{e}\left(x;\zeta_{k|k}^{i,2,q}\right)\label{eq:new_Bernoulli_update_Gaussian}\\
\left(\zeta_{k|k}^{i,2,q},\ell_{2,k|k}^{i,2,q}\right) & =\mathrm{u}_{e}\left(\zeta_{k|k-1}^{e,q},Z_{k}^{j}\right)\\
w_{k|k}^{i,2} & =\delta_{1}\left[|Z_{k}^{j}|\right]\left[\prod_{z\in Z_{k}^{j}}\lambda^{C}\left(z\right)\right]+l_{k|k}^{Z_{k}^{j}}
\end{align}
where $w_{1}^{q}\propto w_{k|k-1}^{p,q}\ell_{1,k|k}^{i,2,q}$ and
$w_{2}^{q}\propto w_{k|k-1}^{e,q}\ell_{2,k|k}^{i,2,q}$. 

For $m_{k}^{j}>1$, 
\begin{align}
l_{k|k}^{Z_{k}^{j}} & =p_{2}^{D}\sum_{q=1}^{n_{k|k-1}^{e}}w_{k|k-1}^{e,q}\ell_{2,k|k}^{i,2,q},
\end{align}
$r_{k|k}^{i,2}=1$ and $c_{k|k}^{i,2}=0$. For $m_{k}^{j}=1$, $Z_{k}^{j}=\left\{ z\right\} $,
we have
\begin{align}
\left(\overline{x}_{k|k}^{i,2,q},P_{k|k}^{i,2,q},\ell_{1,k|k}^{i,2,q}\right) & =\mathrm{u}_{p}\left(\overline{x}_{k|k-1}^{p,q,1},P_{k|k-1}^{p,q,1},z\right)\\
l_{k|k}^{Z_{k}^{j}} & =p_{1}^{D}\sum_{q=1}^{n_{k|k-1}^{p}}w_{k|k-1}^{p,q}\ell_{1,k|k}^{i,2,q}\nonumber \\
 & +p_{2}^{D}\sum_{q=1}^{n_{k|k-1}^{e}}w_{k|k-1}^{e,q}\ell_{2,k|k}^{i,2,q}\\
r_{k|k}^{i,2} & =\frac{l_{k|k}^{Z_{k}^{j}}}{w_{k|k}^{i,a^{i}}}\\
c_{k|k}^{i,2} & =\frac{p_{1}^{D}\sum_{q=1}^{n_{k|k-1}^{p}}w_{k|k-1}^{p,q}\ell_{1,k|k}^{i,2,q}}{l_{k|k}^{Z_{k}^{j}}}.\quad\square
\end{align}
\end{lem}
Lemma \ref{lem:PMBM_update_Gaussian} is obtained by using Theorem
\ref{thm:PMBM_update} and the GGIW and Gaussian updates \cite{Granstrom20,Sarkka_book13}.
We can see that the number of components in the PPP corresponding
to extended targets doubles in the update. This is due to the fact
that the likelihood for misdetection for extended targets, see \eqref{eq:standard_extended_measurement},
has two terms $1-p_{2}^{D}$ and $p_{2}^{D}e^{-\gamma}$. The first
term corresponds to a misdetection obtained through the detection
process modelled by $p_{2}^{D}$, whereas the second term corresponds
to a misdetection obtained when the detection PPP generates zero measurements
\cite{Granstrom12b,Lundquist13,Granstrom20}. These terms create two
updated PPP components for each prior PPP component. For the same
reason, in the update of previous Bernoullis with a misdetection,
the extended target updated density is a mixture of two GGIW, see
\eqref{eq:missed_Bernoulli_update_Gaussian}. As only the Gamma distribution
differs in the two updated GGIWs, we apply merging for Gamma densities
\cite{Granstrom12c} to obtain an updated single-target density of
the form \eqref{eq:single_target_joint_Gaussian}. 

For the detection of previous Bernoullis, the hypothesis represents
with probability $r_{k|k}^{i,a^{i}}=1$ that there is target. If $m_{k}^{j}>1$,
the target is an extended target with probability one ($c_{k|k}^{i,a^{i}}=0$).
If $m_{k}^{j}=1$, the target may be a point or an extended target.
For the new Bernoulli components, if $m_{k}^{j}>1$, the local hypotheses
represent an existing extended target with probability one. For $m_{k}^{j}=1$,
the new Bernoulli may represent clutter, a single target or an extended
target. All possible clutter events are accounted for in the hypotheses
with $m_{k}^{j}=1$ and so do not need to be duplicated in events
with $m_{k}^{j}>1$. We can also see that the single target density
\eqref{eq:new_Bernoulli_update_Gaussian} for new Bernoulli components
is a mixture for both point and extended targets. To obtain an updated
density as in \eqref{eq:single_target_joint_Gaussian}, we perform
merging of the Gaussian mixtures and merging of the GGIW mixtures
\cite{Granstrom12c,Granstrom12d}. 

It should be noted that, if the probability of detection is non-constant,
it can be approximated as a constant at the predicted means for point
and extended targets for each hypothesis \cite[Tab. IV]{Granstrom20}.
Then, we can perform the corresponding updates in Lemma \ref{lem:PMBM_update_Gaussian}.

\subsection{Prediction\label{subsec:Prediction_coexisting}}

We consider that the probability of survival is a constant $p^{S}\left(\cdot\right)=p^{S}$
and linear/Gaussian dynamics for point targets. That is, for $x\in\mathbb{R}^{n_{x}}$,
we have
\begin{align}
g\left(\cdot\left|x\right.\right) & =\mathcal{N}\left(\cdot;Fx,Q\right)
\end{align}
where $F$ is the transition matrix and $Q$ is the process noise
covariance matrix.  For GGIW targets, there are several dynamic models
\cite{Koch08,Granstrom17}. In the simulations, we use the one in
\cite{Granstrom20}. We also assume that a point target cannot become
an extended target and vice versa. The target birth intensity is of
the form
\begin{align}
\lambda_{k}^{B}\left(x\right) & =\sum_{q=1}^{n_{k}^{b,p}}w_{k}^{b,p,q}\mathcal{\mathcal{N}}_{p}\left(x;\overline{x}_{k}^{b,p,q,1},P_{k}^{b,p,q,1}\right)\nonumber \\
 & \quad+\sum_{q=1}^{n_{k}^{b,e}}w_{k}^{b,e,q}\mathcal{G}_{e}\left(x;\zeta_{k}^{b,e,q}\right).\label{eq:PPP_joint_birth}
\end{align}

We apply the PMBM prediction step \cite{Williams15b,Angel18_b} to
obtain a PMBM with the following parameters. Given a single-target
filtering density $f_{k-1|k-1}^{i,a^{i}}\left(\cdot\right)$ of the
form \eqref{eq:single_target_joint_Gaussian}, then the predicted
density is of the same form with $c_{k|k-1}^{i,a^{i}}=c_{k-1|k-1}^{i,a^{i}}$
and
\begin{align}
\zeta_{k|k-1}^{i,a^{i}} & =\mathrm{p}_{e}\left(\zeta_{k-1|k-1}^{i,a^{i}}\right)\\
\left(\overline{x}_{k|k-1}^{i,a^{i},1},P_{k|k-1}^{i,a^{i},1}\right) & =\mathrm{p}_{p}\left(\overline{x}_{k-1|k-1}^{i,a^{i},1},P_{k-1|k-1}^{i,a^{i},1}\right)
\end{align}
where $\mathrm{p}_{p}\left(\cdot\right)$ and $\mathrm{p}_{e}\left(\cdot\right)$
denote the Kalman filter \cite{Sarkka_book13} and the extended target
GGIW prediction \cite[Tab. III]{Granstrom20}, respectively.

The predicted PPP is
\begin{align*}
\lambda_{k|k-1}\left(x\right) & =\sum_{q=1}^{n_{k-1|k-1}^{p}}p^{S}w_{k-1|k-1}^{p,q}\mathcal{\mathcal{N}}_{p}\left(x;\overline{x}_{k|k-1}^{p,q,1},P_{k|k-1}^{p,q,1}\right)\\
 & +\sum_{q=1}^{n_{k-1|k-1}^{e}}p^{S}w_{k-1|k-1}^{e,q}\mathcal{G}_{e}\left(x;\zeta_{k|k-1}^{e,q}\right)+\lambda_{k}^{B}\left(x\right)
\end{align*}
where $\left(\overline{x}_{k|k-1}^{p,q,1},P_{k|k-1}^{p,q,1}\right)=\mathrm{p}_{p}\left(\overline{x}_{k-1|k-1}^{p,q,1},P_{k-1|k-1}^{p,q,1}\right)$
and $\zeta_{k|k-1}^{e,q}=\mathrm{p}_{e}\left(\zeta_{k-1|k-1}^{e,q}\right)$.

While this prediction step assumes that there is no dynamic change
between point and extended targets (e.g., the targets are pedestrians
and vehicles), in some applications, a point target may become an
extended target if it gets sufficiently close to the sensor. In this
setting, one should design the corresponding transition density to
capture this.

\subsection{PMB approximation\label{subsec:PMB-approximation}}

It is also useful to consider a PMB approximation to the PMBM \eqref{eq:PMBM}
to develop a faster algorithm. If we perform this approximation after
each update, we obtain the corresponding PMB filter \cite{Williams15b,Williams15}.
Given an updated PMBM \eqref{eq:PMBM} with $k'=k$, the track-oriented
PMB approximation is
\begin{align}
f_{k|k}^{\mathrm{pmb}}\left(X_{k}\right) & =\sum_{Y\uplus W=X_{k}}f_{k|k}^{\mathrm{p}}\left(Y\right)f_{k|k}^{\mathrm{mb}}\left(W\right)\label{eq:PMB_1}\\
f_{k|k}^{\mathrm{mb}}\left(X_{k}\right) & =\sum_{\uplus_{l=1}^{n_{k|k}}X^{l}=X_{k}}\prod_{i=1}^{n_{k|k}}f_{k|k}^{i}\left(X^{i}\right)\label{eq:PMB_2}
\end{align}
where $f_{k|k}^{i}\left(\cdot\right)$ is a Bernoulli density with
probability $r^{i}$ of existence and single target density $p^{i}\left(\cdot\right)$
such that
\begin{align}
r^{i} & =\sum_{a^{i}=1}^{h^{i}}\overline{w}_{k|k}^{i,a^{i}}r_{k|k}^{i,a^{i}}\label{eq:existence_Prop2}\\
p^{i}\left(x\right) & =\frac{\sum_{a^{i}=1}^{h^{i}}\overline{w}_{k|k}^{i,a^{i}}r_{k|k}^{i,a^{i}}f_{k|k}^{i,a^{i}}\left(x\right)}{r^{i}}\label{eq:p_i_Prop2}\\
\overline{w}_{k|k}^{i,a^{i}} & =\sum_{b\in\mathcal{A}_{k|k}:b^{i}=a^{i}}w_{k|k}^{b}.
\end{align}
The PMB approximation \eqref{eq:PMB_1}-\eqref{eq:PMB_2} minimises
the Kullback-Leibler divergence (KLD) on a single target space augmented
with an auxiliary variable, which represents if the target remains
undetected or corresponds to the $i$-th Bernoulli component \cite{Angel20_e}.
We can see that \eqref{eq:p_i_Prop2} is a mixture over all local
hypotheses and that the PPP part of the PMBM \eqref{eq:PMBM} is not
affected by the PMB approximation.

In the implementation for coexisting point-extended targets, we are
interested in single target densities of the form \eqref{eq:single_target_joint_Gaussian}.
By using moment matching (KLD minimisation) for the mixture in $p^{i}\left(\cdot\right)$,
we obtain the single-target density

\begin{align}
p^{i}\left(x\right) & =c^{i}\mathcal{N}_{p}\left(x;\overline{x}_{k|k}^{i},P_{k|k}^{i}\right)\nonumber \\
 & \quad+\left(1-c^{i}\right)\mathcal{G}_{e}\left(x;\zeta_{k|k}^{i}\right)\\
c^{i} & =\frac{\sum_{a^{i}=1}^{h^{i}}\overline{w}_{k|k}^{i,a^{i}}r_{k|k}^{i,a^{i}}c_{k|k}^{i,a^{i}}}{r^{i}}\\
\left(\overline{x}_{k|k}^{i},P_{k|k}^{i}\right) & =\mathrm{m}_{\mathrm{G}}\left(\overline{x}_{k|k}^{i,a^{1},1},P_{k|k}^{i,a^{1},1},...,,\overline{x}_{k|k}^{i,a^{h^{i}},1},P_{k|k}^{i,a^{h^{i}},1},\right.\nonumber \\
 & \left.\beta_{\mathrm{G}}^{i,a^{1}},...,\beta_{\mathrm{G}}^{i,a^{h^{i}}}\right)\\
\beta_{\mathrm{G}}^{i,a^{i}} & \propto\overline{w}_{k|k}^{i,a^{i}}r_{k|k}^{i,a^{i}}c_{k|k}^{i,a^{i}}\\
\zeta_{k|k}^{i} & =\mathrm{m}_{\mathrm{GG}}\left(\zeta_{k|k}^{i,a^{1}},...,\zeta_{k|k}^{i,a^{h^{i}}},\beta_{\mathrm{GG}}^{i,a^{1}},...,\beta_{\mathrm{GG}}^{i,a^{h^{i}}}\right)\\
\beta_{\mathrm{GG}}^{i,a^{i}} & \propto\overline{w}_{k|k}^{i,a^{i}}r_{k|k}^{i,a^{i}}\left(1-c_{k|k}^{i,a^{i}}\right)
\end{align}
where $\mathrm{m}_{\mathrm{G}}\left(\cdot\right)$ is a function that
obtains the mean and covariance of a Gaussian mixture with weights
$\beta_{\mathrm{G}}^{i,a^{1}},...,\beta_{\mathrm{G}}^{i,a^{h^{i}}}$
(normalised to sum to one) and moments $\overline{x}_{k|k}^{i,a^{1},1},P_{k|k}^{i,a^{1},1},...,,\overline{x}_{k|k}^{i,a^{h^{i}},1},P_{k|k}^{i,a^{h^{i}},1}$
\cite{Bishop_book06}. The function $\mathrm{m}_{\mathrm{GG}}\left(\cdot\right)$
obtains the GGIW parameters that minimise the KLD from a mixture with
weights $\beta_{\mathrm{GG}}^{i,a^{1}},...,\beta_{\mathrm{GG}}^{i,a^{h^{i}}}$
(normalised to sum to one) and parameters $\zeta_{k|k}^{i,a^{1}},...,\zeta_{k|k}^{i,a^{h^{i}}}$
\cite{Granstrom12d,Granstrom12c}. 

\subsection{Target state estimation\label{subsec:Estimation}}

Given a PMBM posterior, we can apply several estimators to estimate
the current set of targets, see details in \cite[Sec. VI]{Angel18_b}.
We proceed to explain how Estimator 1 in \cite[Sec. VI]{Angel18_b},
which is the one we use in the simulations, is adapted to deal with
the single-target space $\mathcal{X}=\mathbb{R}^{n_{x}}\uplus\mathcal{X}_{e}$. 

We first obtain the global hypothesis with highest weight and select
its Bernoulli components whose probability of existence is above a
threshold (0.5 in the simulations). For each of these Bernoulli components,
which have densities of the form \eqref{eq:single_target_joint_Gaussian},
we estimate a target state, which may be a point or an extended target.
If the probability of being a point target is $c_{k|k}^{i,a^{i}}>0.5$,
then we estimate a point target located at the mean $\overline{x}_{k|k}^{i,a^{i},1}$.
Otherwise, we estimate an extended target with kinematic and extent
states located at the mean \cite{Koch08}
\begin{align}
\hat{\xi}_{k} & =\overline{x}_{k|k}^{i,a^{i},2}\\
\hat{X}_{k} & =\frac{V_{k|k}^{i,a^{i}}}{v_{k|k}^{i,a^{i}}-2d-2}.
\end{align}

\subsection{Practical aspects\label{subsec:Practical-aspects}}

As in other multiple target filters with data associations, the number
of global and local hypotheses increases unboundedly in time. Therefore,
in practice, it is necessary to perform approximations, with the objective
of only propagating hypotheses with relevant weights. In fact, due
to the structure of the hypotheses of Theorem \ref{thm:PMBM_update},
the way to handle the data association problem with coexisting point
and extended targets is quite similar to the extended target case
\cite{Granstrom17,Granstrom20}. 

In our implementation, the PMBM posterior is represented by a list
of Bernoullis $i\in\left\{ 1,...,n_{k|k'}\right\} $, where each of
them contains their local hypotheses with their parameters, a global
hypothesis table, which contains indices to local hypotheses of each
Bernoulli, and a vector with the global hypotheses weights. To deal
with the data association problem at each update, we first perform
gating to obtain two sets of measurements: 1) measurements that are
in the gate of at least one previous Bernoulli, and 2) measurements
that are only in the gate of the PPP components. Measurements that
do not fall into these categories are discarded. 

For the set of measurements in group 1), we first generate possible
partitions of this set using the DBSCAN algorithm with distance thresholds
between $\Gamma_{d,min}$ and $\Gamma_{d,max}$, with a step size
of $\varepsilon_{d}$ \cite{Ester96,Xia_arxiv18}. The minimum number
of points to form a region, which is a parameter of the DBSCAN algorithm,
is set to 1 to capture point-target measurements. Among the possible
partitions generated by the multiple runs of DBSCAN algorithms, there
may be repeated ones, so we keep the unique ones and we obtain the
unique subsets of measurements in these partitions. These subsets
are then used to generate the updated local hypotheses for previous
Bernoullis, see \eqref{eq:Miss_measurement}-\eqref{eq:update_density}.
A new Bernoulli component is also created for each unique subset of
measurements that is in the gate of a GGIW PPP component. For each
previous global hypothesis and partition, obtained by DBSCAN, we run
Murty's algorithm \cite{Murty68} to find the global hypotheses with
highest weights.

For the set of measurements in group 2), which may correspond to newly
detected targets, we run the DBSCAN to obtain possible partitions.
Each of these partitions in theory gives rise to different global
hypotheses corresponding to new born targets. We simplify this procedure
by finding the partition with highest weight and only generating the
Bernoulli components that are generated by the sets in this partition
\cite{Granstrom20}. These new Bernoulli components are added to all
the global hypotheses, whose weights remain unchanged.

We would like to point out that, while DBSCAN is a fast method for
clustering, it is agnostic to target shape. Therefore, in difficult
scenarios, it may be suitable to consider further partitions using
additional methods that account for target shape, for example, prediction
partition and expectation maximisation partition \cite{Granstrom12b,Granstrom18b}.

We also perform pruning of global hypotheses with low weights, and
pruning of Bernoulli components with low existence probabilities \cite{Angel19_e,Angel18_b,Angel20}.
A pseudocode of the resulting PMBM update is provided in Algorithm
\ref{alg:Pseudocode_update}. The  PMB filter performs the same PMBM
update and it is then followed by the PMB approximation, see Section
\ref{subsec:PMB-approximation}. It is also possible to approximate
the PMB marginal data association probabilities directly using belief
propagation \cite{Meyer20b,Meyer20,Meyer18}. 

\begin{algorithm}
\caption{\label{alg:Pseudocode_update}Pseudocode of the PMBM update}

{\fontsize{9}{9}\selectfont

\begin{algorithmic}     

\State - Perform gating to separate current measurements into the
following disjoint categories: 

\State $\quad\circ$ 1. A set of measurements that are in the gate
of at least one previous Bernoulli.

\State $\quad\circ$ 2. A set of measurements that are only in the
gate of PPP components.

\State - For measurements corresponding to 1:

\State $\quad\circ$ Run DBSCAN to generate possible partitions.

\State $\quad\circ$ Obtain unique subsets in the previous partitions. 

\State $\quad\circ$ Generate new local hypotheses for previous Bernoullis.

\State $\quad\circ$ Generate new Bernoulli components.

\State $\quad\circ$ For each previous global hypothesis, run Murty's
algorithm to obtain updated global hypotheses.

\State - Perform pruning of global hypotheses and Bernoulli components.

\State - For measurements corresponding to 2 (new targets):

\State $\quad\circ$ Run DBSCAN to generate possible partitions.

\State $\quad\circ$ Find the partition with highest weight.

\State $\quad\circ$ Generate the new Bernoulli components for this
partition.

\State $\quad\circ$ Add these Bernoullis to the global hypotheses.

\end{algorithmic}

}
\end{algorithm}

\section{Simulations}

\label{sec:Simulations}

In this section, we assess the PMBM and PMB filters for coexisting
point and extended targets via numerical simulations\footnote{Matlab code is available at https://github.com/Agarciafernandez and
https://github.com/yuhsuansia.}. In this section, we refer to these filters as point-extended PMBM
and PMB (PE-PMBM and PE-PMB) filters. The filters are implemented
with the following parameters: maximum number of hypotheses $20$,
threshold for pruning the PPP weights $10^{-5}$, threshold for pruning
Bernoulli components $10^{-3}$ and threshold for pruning global hypotheses
$10^{-3}$. The DBSCAN algorithm \cite{Ester96} is run with distance
thresholds between $\Gamma_{d,min}=0.1$ and $\Gamma_{d,max}=12$,
with a step size of $\varepsilon_{d}=0.1$.  We have also implemented
a point-extended MBM (PE-MBM) filter, see Section \ref{subsec:Discussion}.

Extended target filters can in principle deal with point-target detections,
as they do not place zero probability to this event. Therefore, we
compare the proposed filters with extended target PMBM and PMB filters,
which we refer to as E-PMBM and E-PMB filters \cite{Granstrom20,Xia_arxiv18}.
We proceed to discuss the models and the simulations results. All
the units in this section are given in the international system.

\subsection{Models}

We consider a point target state $\left[p_{x},\dot{p}_{x},p_{y},\dot{p}_{y}\right]^{T}$,
which contains position and velocity in a two-dimensional plane. Point
targets move with a nearly-constant velocity model with 
\begin{align*}
F=I_{2}\otimes\left(\begin{array}{cc}
1 & \tau\\
0 & 1
\end{array}\right),\quad Q=qI_{2}\otimes\left(\begin{array}{cc}
\tau^{3}/3 & \tau^{2}/2\\
\tau^{2}/2 & \tau
\end{array}\right)
\end{align*}
where $\tau=1$, $q=0.25$, $\otimes$ denotes Kronecker product and
$I_{2}$ is an identity matrix of size 2. The probability of survival
is $p^{S}=0.99$. The extended target model is the GGIW model in Section
\ref{sec:Coexisting_point_extended}. Extended targets move with the
previous nearly-constant velocity model and their extent matrix and
$\gamma$ parameter remain constant. The probability of survival is
0.99. 

The birth model is a PPP of the form \eqref{eq:PPP_joint_birth}.
The PPP point target part has parameters $n_{k}^{b,p}=1$, $w_{k}^{b,p,1}=0.03$,
$\overline{x}_{k}^{b,p,1,1}=\left[0,0,0,0\right]^{T}$, $P_{k}^{b,p,1,1}=\mathrm{diag}([200^{2},4^{2},200^{2},4^{2}])$.
The extended target part has: $n_{k}^{b,e}=1$, $w_{k}^{b,e,1}=0.06$,
and
\begin{align*}
\zeta_{k}^{b,e,q} & =\left(40,4,\overline{x}_{k}^{b,p,1,1},P_{k}^{b,p,1,1},20,200I_{2}\right).
\end{align*}
As the birth covariance matrix is large, new born targets may appear
in a large area. The multi-Bernoulli birth model for the PE-MBM filter
has a single Bernoulli with existence probability 0.06, point-target
probability $c=1/3$, point target mean $\overline{x}_{k}^{b,p,1,1}$
and covariance $P_{k}^{b,p,1,1}$, and GGIW $\zeta_{k}^{b,e,q}$. 

\begin{figure}
\begin{centering}
\includegraphics[scale=0.6]{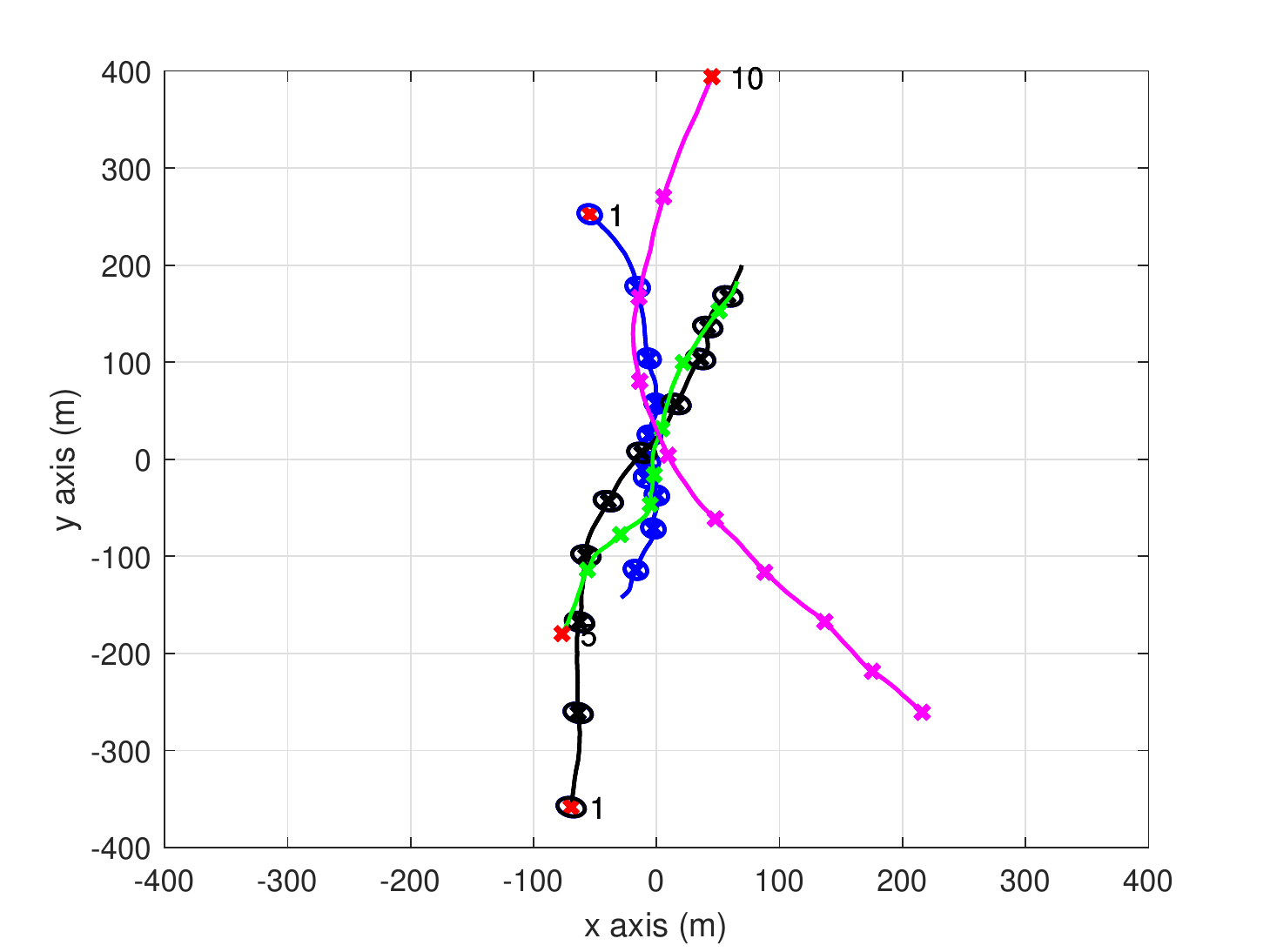}
\par\end{centering}
\caption{\label{fig:Scenario}Scenario of the simulations. Two extended targets
are born at time step 1 and two point targets are born at time steps
5 and 10. The extended targets are alive at all time steps. The last
time steps of the point targets are 38 and 60. Targets are in close
proximity at around time step 50. Target states at time of birth are
marked with a red cross, and every 10 time steps with a cross. The
3-$\sigma$ ellipse for extended targets is shown every 10 time steps. }
\end{figure}

We measure the positions of the targets. For point targets, we have
parameters: $p_{1}^{D}=0.95$, and
\begin{align*}
H_{1}=\left(\begin{array}{cccc}
1 & 0 & 0 & 0\\
0 & 0 & 1 & 0
\end{array}\right), & \quad R=\sigma^{2}I_{2}
\end{align*}
where $\sigma^{2}=1$. For extended targets, the parameters are $p_{2}^{D}=p_{1}^{D}$,
$H_{2}=H_{1}.$ Clutter is uniformly distributed in the surveillance
area $\left[-500,500\right]\times\left[-500,500\right]$ with an average
of $\lambda^{C}=8$ false alarms per scan. We consider 100 time steps
and the set of trajectories shown in Figure \ref{fig:Scenario}, which
has been obtained by sampling from the dynamic process. The E-PMBM
and E-PMB filters are recovered by setting the birth intensity for
point targets to zero in the PE-PMBM and PE-PMB filters. 

\subsection{Results}

\begin{figure}
\begin{centering}
\includegraphics[scale=0.6]{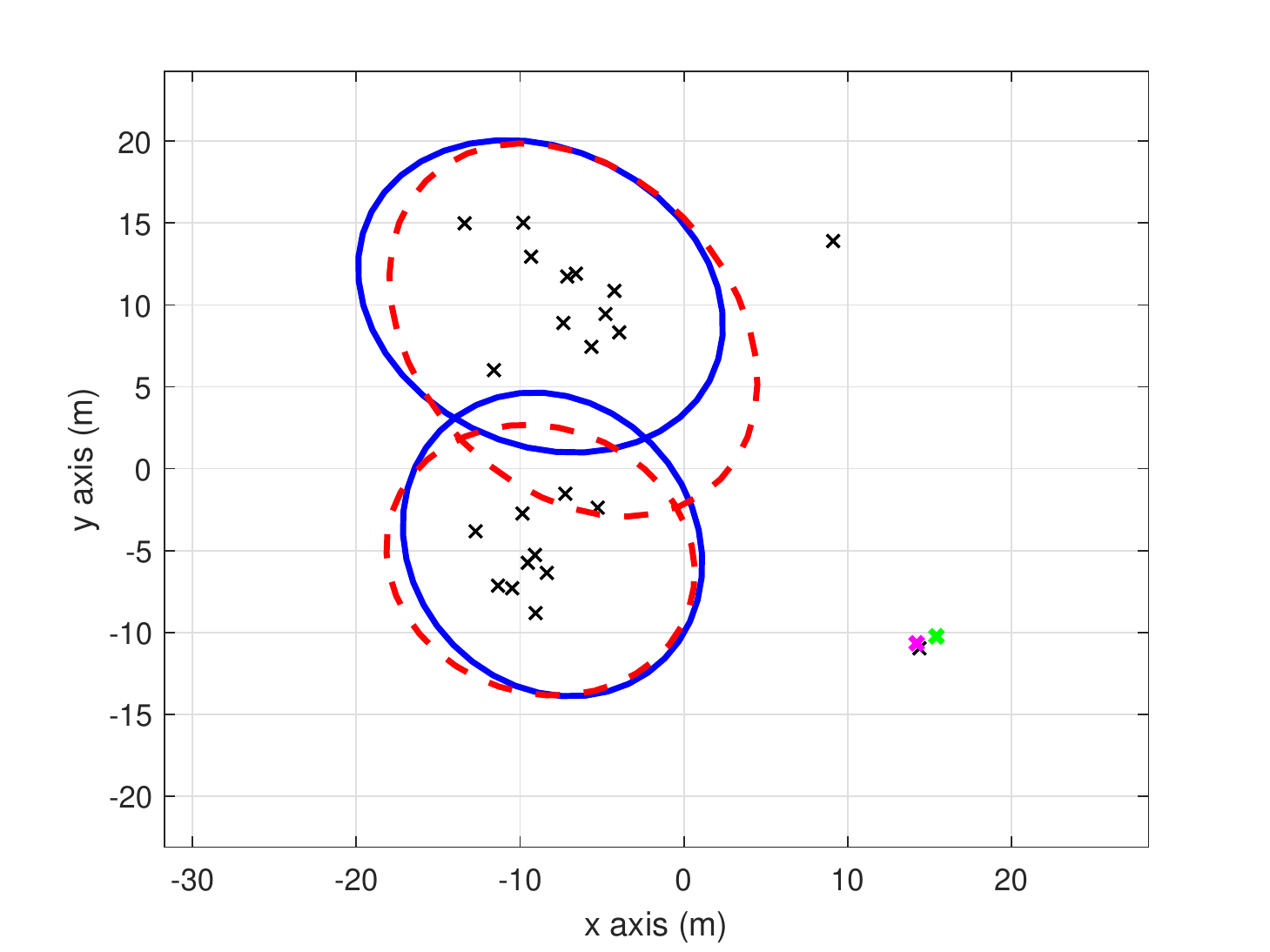}
\par\end{centering}
\caption{\label{fig:Estimate_k52} Ground truth and estimated set of targets
at time step 52 in an illustrative run. Measurements are shown as
black crosses. Blue ellipses represent the true extended targets,
a green cross represents the true point target. The red, dashed ellipses
represent the estimated extended targets and the pink cross represents
the estimated point target. The three targets are properly detected
and classified. At this time step, there are 22 measurements within
the gate of the previous targets (the three targets shown in the figure).
The output of the DBSCAN algorithm with different distance thresholds
produces 18 partitions of these measurements, ranging from partitions
with 22 clusters (each with a single measurement) to 2 clusters.}
\end{figure}

We first show the ground truth and the estimate of the set of targets
at time step 52 in an illustrative run with the PE-PMBM filter in
Figure \ref{fig:Estimate_k52}. We can see that the each extended
target generates several measurements and are detected. The ellipses
of the estimated targets are reasonably accurate. The point target
generates a single measurement at this time step, and it is also detected.
Its estimate is close to its true state. In this scenario, the class
probability quickly reaches either zero or one for the considered
targets, classifying all targets correctly. 

We evaluate filter performance via Monte Carlo simulation with 100
runs. We compute the error between the true set of targets at each
time and its estimate using the generalised optimal subpattern assignment
(GOSPA) metric with parameters $\alpha=2$, $p=2$, $c=10$, and its
decomposition into localisation errors and costs for missed and false
targets \cite{Rahmathullah17}. The base metric for target states
is the Gaussian Wasserstein distance, which measures error for position
and extent \cite{Yang16}. In the base metric, we consider a point
target as an extended target with extent zero.

\begin{figure}
\begin{centering}
\includegraphics[scale=0.3]{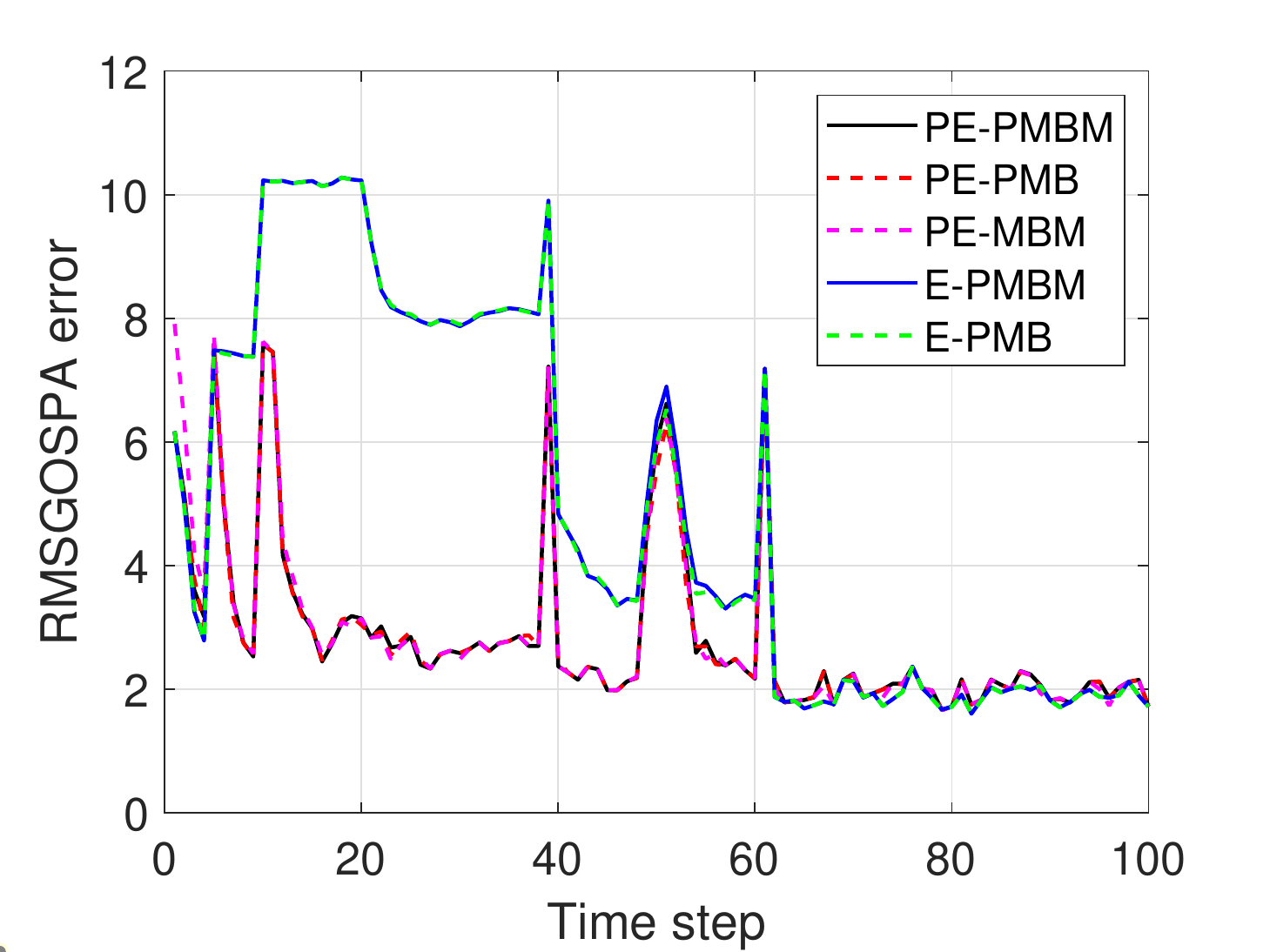}\includegraphics[scale=0.3]{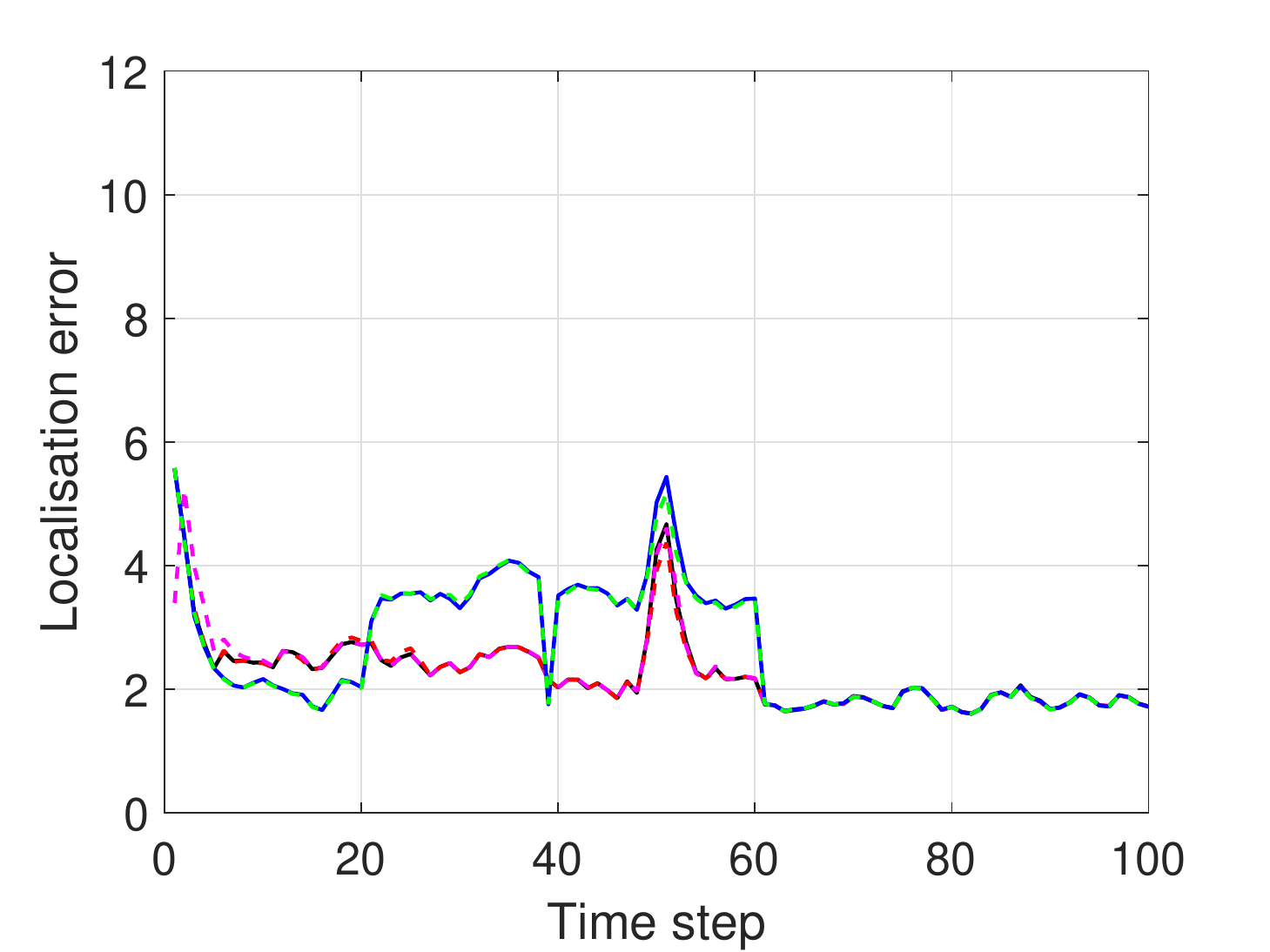}
\par\end{centering}
\begin{centering}
\includegraphics[scale=0.3]{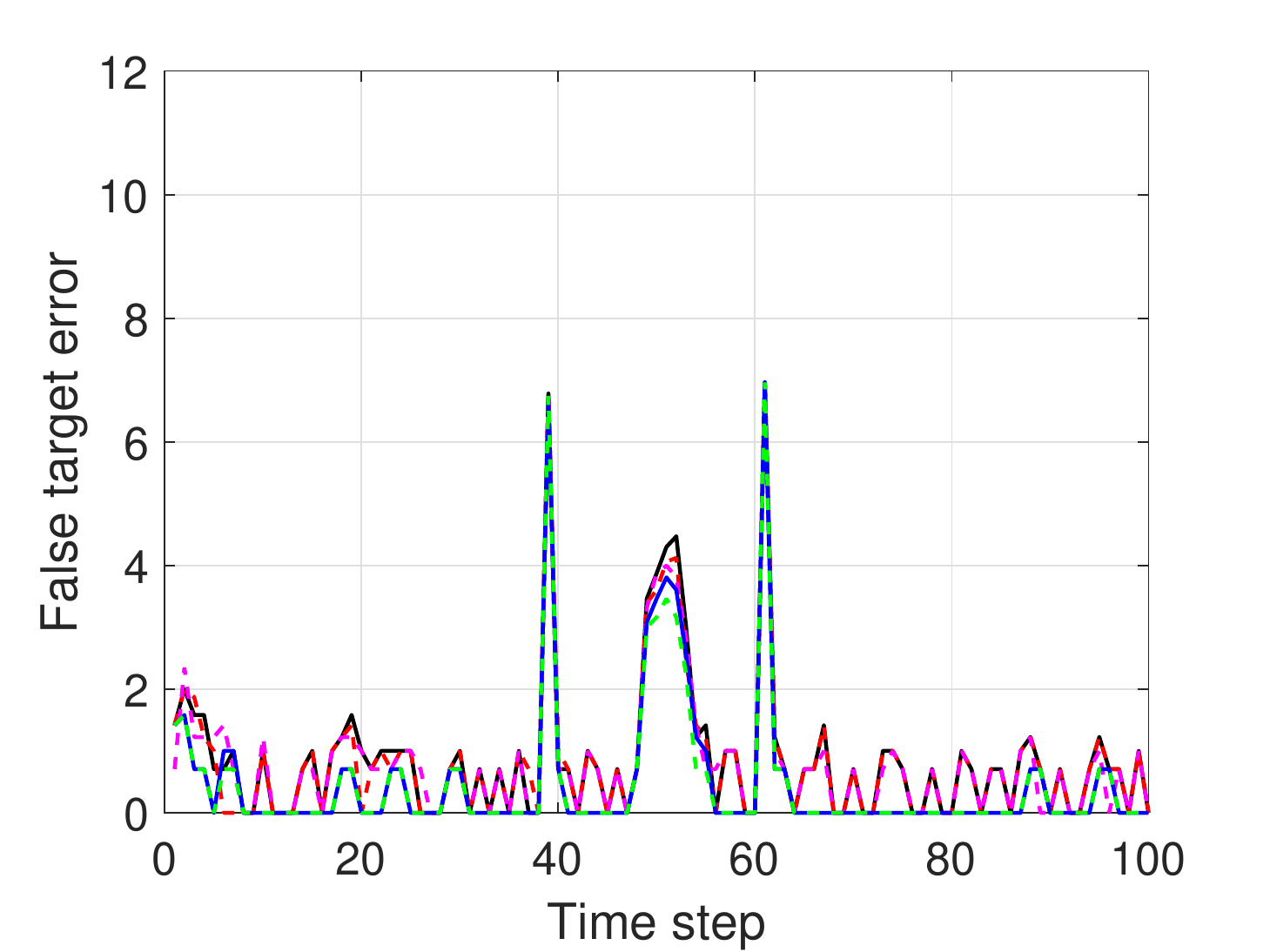}\includegraphics[scale=0.3]{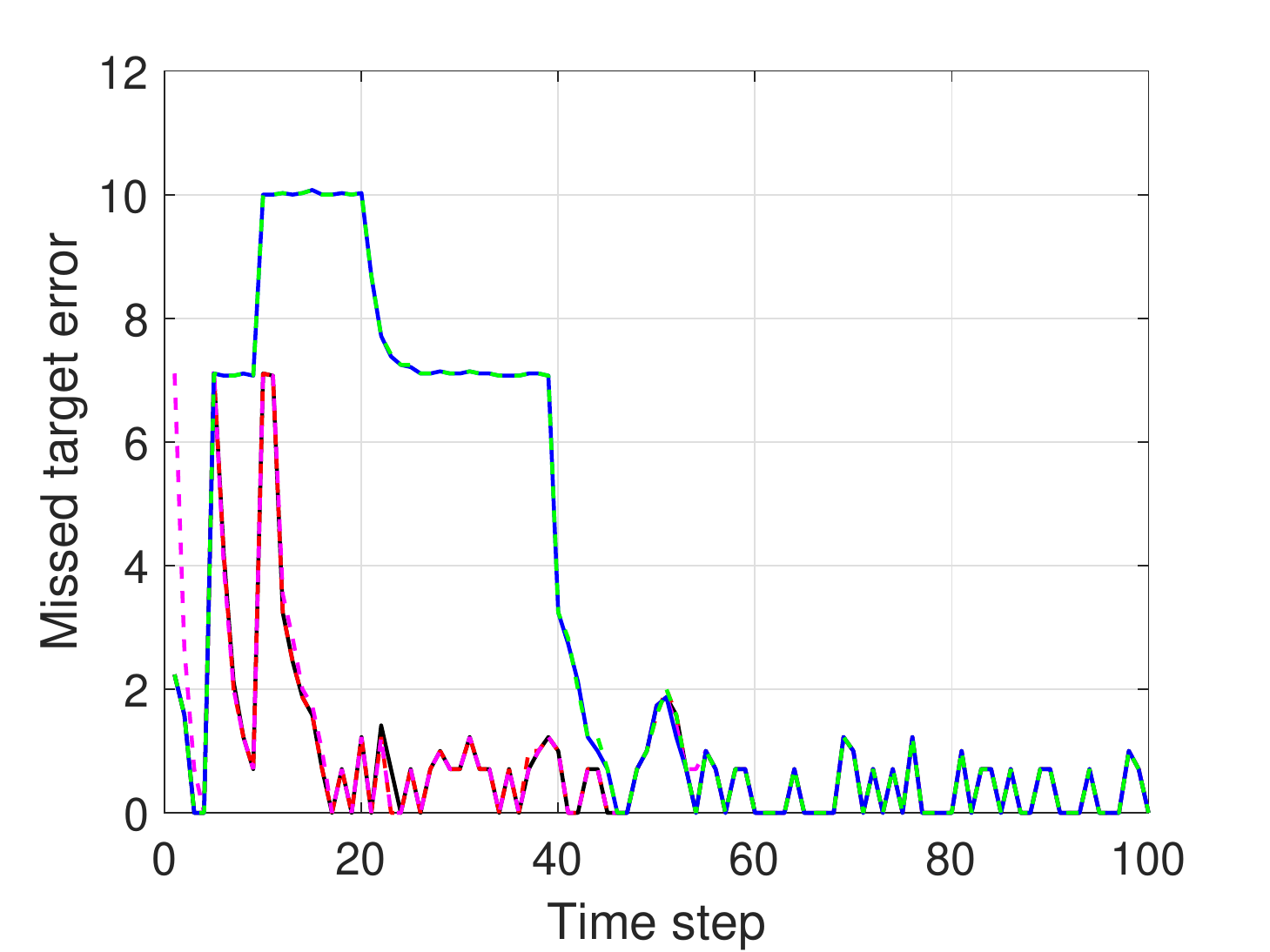}
\par\end{centering}
\caption{\label{fig:RMS-GOSPA-error}RMS-GOSPA error (m) for the position elements
and its decomposition. The PE-PMBM and PE-PMB filters have very similar
performance. PE-MBM fails to detect one of the targets at time step
1. The E-PMBM and E-PMB have a higher error at some time steps due
to missed point targets. E-PMBM and E-PMB localisation errors are
also higher at some time steps, as point targets are estimated with
some extent.}
\end{figure}

The root mean square GOSPA (RMS-GOSPA) error against time and its
decomposition are shown in Figure \ref{fig:RMS-GOSPA-error}. We can
see that PE-PMBM and PE-PMB filter perform quite similarly and outperform
E-PMBM and E-PMB. PE-MBM performs quite similarly to PE-PMBM and PE-PMB
but does not detect one of the targets at time step 1, as the birth
model sets the maximum number of new born targets to one. For PE-PMBM
and PE-PMB, missed target errors are higher when new targets are born.
False target errors are higher when targets die and when targets get
in close proximity. Localisation errors are higher at the beginning
of the simulation, and when targets get in close proximity, as the
data association problem is more complicated. ET-PMBM and ET-PMB also
behave quite similarly and have more difficulty in detecting the point
targets, so they show a higher missed target error at some time steps.
In addition, the localisation error is also higher at some time steps,
as point targets are estimated with a certain extent, which increases
the error compared to the ground truth. 

The running times of the Matlab implementations (100 time steps) on
an Intel Core i5 laptop are: 56.4s (PE-PMBM), 17.5s (PE-PMB), 64.5
(PE-MBM), 25.5s (E-PMBM) and 15.2 (E-PMB). The PMB filters are considerably
faster than PMBM/MBM, as they do not propagate a mixture through the
filtering recursion. Only considering extended targets is also faster,
though it decreases performance.

To provide more complete simulation results, we show the RMS-GOSPA
errors, along with the GOSPA error decomposition, considering all
time steps for different values of the probability of detection and
clutter rate in Table \ref{tab:GOSPA_different_parameters}. Due to
space constraints, we do not show E-PMB, which behaves quite similarly
to E-PMBM. In this table, ``Tot.'', ``Loc.'', ``Fal.'' and ``Mis.''
refer to total GOSPA, localisation, false target and missed target
costs, respectively. The filters with coexisting point extended targets
consistently provide more accurate results, especially due to a lower
number of missed targets. The PE-PMBM and PE-PMB filters provide quite
similar results though the PE-PMB filter is slightly better. While
the PE-PMBM filter provides the closed-form solution to the filtering
recursion, we apply approximations and an suboptimal estimator, so
the PE-PMB filter may work better in some scenarios. Decreasing the
probability of detection or increasing the clutter rate, the GOSPA
error for all filters increases, mainly due to a rise in missed target
cost.

\begin{table*}
\caption{\label{tab:GOSPA_different_parameters}RMS-GOSPA errors and their
decompositions for the filters and different parameters}

\centering{}%
\begin{tabular}{c|c|cccc|cccc|cccc||cccc}
\hline 
 &
 &
\multicolumn{4}{c|}{PE-PMBM} &
\multicolumn{4}{c|}{PE-PMB} &
\multicolumn{4}{c||}{PE-MBM} &
\multicolumn{4}{c}{E-PMBM}\tabularnewline
\cline{3-18} \cline{4-18} \cline{5-18} \cline{6-18} \cline{7-18} \cline{8-18} \cline{9-18} \cline{10-18} \cline{11-18} \cline{12-18} \cline{13-18} \cline{14-18} \cline{15-18} \cline{16-18} \cline{17-18} \cline{18-18} 
$p_{1}^{D}=p_{2}^{D}$ &
$\lambda_{c}$ &
Tot. &
Loc. &
Fal. &
Mis. &
Tot. &
Loc. &
Fal. &
Mis. &
Tot. &
Loc. &
Fal. &
Mis. &
Tot. &
Loc. &
Fal. &
Mis.\tabularnewline
\hline 
$0.95$ &
$8$ &
3.21 &
2.36 &
1.50 &
1.57 &
\uline{3.18} &
2.35 &
1.46 &
1.56 &
3.27 &
2.36 &
1.45 &
1.73 &
5.81 &
2.83 &
1.29 &
4.91\tabularnewline
$0.95$ &
$16$ &
3.38 &
2.48 &
1.54 &
1.70 &
\uline{3.35} &
2.46 &
1.51 &
1.70 &
3.49 &
2.45 &
1.42 &
2.04 &
6.91 &
2.14 &
0.86 &
6.52\tabularnewline
$0.85$ &
$8$ &
3.70 &
2.61 &
1.90 &
1.81 &
\uline{3.65} &
2.60 &
1.80 &
1.82 &
3.84 &
2.52 &
1.42 &
2.54 &
6.54 &
2.74 &
1.31 &
5.79\tabularnewline
$0.85$ &
$16$ &
3.78 &
2.60 &
1.86 &
2.02 &
\uline{3.77} &
2.64 &
1.80 &
2.01 &
4.09 &
2.51 &
1.33 &
2.94 &
7.10 &
2.25 &
0.88 &
6.68\tabularnewline
\hline 
\end{tabular}
\end{table*}

\section{Conclusions}

\label{sec:Conclusions}

We have derived the update of a PMBM filter with a measurement model
that can consider point and extended targets, and we have shown that
the updated posterior is also a PMBM. We have also proposed an implementation
of the resulting PMBM recursion to consider coexisting point and extended
targets. In order to do so, we first set the suitable single-target
space and single target densities, which are based on Gaussian densities
for single targets, and GGIW densities for extended targets. Finally,
based on the previous results, we have explained how to obtain a computationally-lighter
PMB filter for coexisting point and extended targets.

We think there are many lines of future work. In many applications,
there are coexisting point and extended targets, and one can perform
research into tailored measurement and target models for each application.
Another line of future work is to extend the above results to consider
PMBMs on sets of trajectories, with coexisting point and extended
targets, to provide full trajectory information \cite{Angel20_b,Xia19_b,Xia19}.

\bibliographystyle{IEEEtran}
\bibliography{6C__Trabajo_laptop_Mis_articulos_Finished_Coexisting_point-extended_PMBM_Accepted_Referencias}

\cleardoublepage{}

{\LARGE{}Supplementary material: A Poisson multi-Bernoulli mixture
filter for coexisting point and extended targets}{\LARGE\par}

\appendices{}

\section{\label{sec:Update_proof_App}}

In this appendix, we prove Theorem \ref{thm:PMBM_update}, which provides
the update step, making use of probability generating functionals
(PGFLs). A PGFL is an alternative representation of a multi-object
density, in the same way as Fourier and $z$-transforms are for signals
defined in the time domain. 

For a multi-object density $f\left(\cdot\right)$, its PGFL $G_{f}[\cdot]$
is given by the set integral \cite{Mahler_book14} 
\begin{align}
G_{f}[h] & =\int h^{X}f(X)\delta X\label{eq:PGLF_definition}
\end{align}
where $h\left(\cdot\right)$ is a unitless function of state space,
and $h^{X}=\prod_{x\in X}h(x)$ , $h^{\emptyset}=1$. The test function
for PGFLs related to densities defined for targets and measurements
are denoted as $h\left(\cdot\right)$ and $g\left(\cdot\right)$,
respectively.

Given the PGFL $G_{f}[\cdot]$, we can recover its multi-object density
$f\left(\cdot\right)$ by the set derivative \cite{Mahler_book14}
\begin{align}
f\left(X\right) & =\frac{\delta}{\delta X}G_{f}[h]\bigg|_{h=0}.\label{eq:inverse_PGFL}
\end{align}

\subsection{PGFLs of targets and measurements}

The density \eqref{eq:PMBM} in PGFL form is represented as \cite{Williams15b}
\begin{align}
G_{k|k'}[h] & =G_{k|k'}^{\mathrm{p}}[h]\cdot G_{k|k'}^{\mathrm{mbm}}[h]\label{eq:Hybrid_PGFL}\\
G_{k|k'}^{\mathrm{p}}[h] & =\exp\left(\langle\lambda_{k|k'},h-1\rangle\right)\propto\exp\left(\langle\lambda_{k|k'},h\rangle\right)\label{eq:PPP_PGFL}\\
G_{k|k'}^{\mathrm{mbm}}[h] & =\sum_{a\in\mathcal{A}_{k'|k}}w_{k|k'}^{a}\prod_{i=1}^{n_{k'|k}}G_{k|k'}^{i,a^{i}}[h]\label{eq:MBM_PGFL}\\
 & \propto\sum_{a\in\mathcal{A}_{k'|k}}\prod_{i=1}^{n_{k|k'}}\left[w_{k|k'}^{i,a^{i}}G_{k|k'}^{i,a^{i}}[h]\right]
\end{align}
where
\begin{align}
G_{k|k'}^{i,a^{i}}[h] & =1-r_{k|k'}^{i,a^{i}}+r_{k|k'}^{i,a^{i}}\langle f_{k|k'}^{i,a^{i}},h\rangle.
\end{align}

Given the multi-target state $X$, measurements from each target are
independent, and there is also independent PPP clutter. Therefore,
the PGFL $G_{Z}[g|X]$ of the measurements given $X$ is the product
of PGFL
\begin{align}
G_{Z}[g|X] & =\exp\left(\langle\lambda^{C},g-1\rangle\right)\prod_{x\in X}G[g|x]
\end{align}
where $G[g|x]$ is the PGFL of $f\left(Z|x\right)$. 

\subsection{Joint PGFL of targets and measurements}

The joint PGFL of measurements and targets is \cite{Williams15b,Mahler_book14}
\begin{align}
F[g,h] & =\int\int g^{Z_{k}}h^{X_{k}}f\left(Z_{k}|X_{k}\right)f_{k|k-1}\left(X_{k}\right)\delta Z_{k}\delta X_{k}\\
 & =\int G[g|X_{k}]h^{X_{k}}f_{k|k-1}\left(X_{k}\right)\delta X_{k}\\
 & =\exp\left(\langle\lambda^{C},g-1\rangle\right)G_{k|k-1}[hG_{Z}[g|\cdot]]\\
 & \propto\exp\left(\langle\lambda^{C},g\rangle+\langle\lambda_{k|k-1},hG[g|\cdot]\rangle\right)\nonumber \\
 & \times\sum_{a\in\mathcal{A}_{k|k-1}}\prod_{i=1}^{n_{k|k-1}}\left[w_{k|k-1}^{i,a^{i}}G_{k|k-1}^{i,a^{i}}\big[hG[g|\cdot]\big]\right].\label{eq:JointPGFLFull}
\end{align}

We denote the first line of \eqref{eq:JointPGFLFull} as
\begin{align}
F^{0}[g,h] & =\exp\left(\langle\lambda^{C},g\rangle+\big\langle\lambda_{k|k-1},hG[g|\cdot]\big\rangle\right)\label{eq:JointPGFLPoissonComponent}
\end{align}
which represents the joint PGFL of measurements (including false alarms)
and targets in the PPP, up to a proportionality constant. We also
denote
\begin{align}
F^{i,a^{i}}[g,h] & =G_{k|k-1}^{i,a^{i}}\big[hG[g|\cdot]\big]\\
 & =1-r_{k|k-1}^{i,a^{i}}+r_{k|k-1}^{i,a^{i}}\bigg\langle f_{k|k-1}^{i,a^{i}},hG[g|\cdot]\bigg\rangle\label{eq:JointPGFLBernoulliComponent}
\end{align}
which represents the joint PGFL of measurements (not including false
alarms) and the $i$-th potential target. Then, using \eqref{eq:global_weights},
we can write \eqref{eq:JointPGFLFull} as
\begin{align}
F[g,h] & \propto F^{0}[g,h]\sum_{a\in\mathcal{A}_{k|k-1}}\prod_{i=1}^{n_{k|k-1}}\left[w_{k|k-1}^{i,a}F^{i,a^{i}}[g,h]\right].
\end{align}

\subsection{Updated PGFL}

We calculate the updated density $f_{k|k}\left(\cdot\right)$ via
its PGFL $G_{k|k}[h]$, which is given by the set derivative of $F[g,h]$
w.r.t. $Z_{k}$ evaluated at $g=0$ \cite[Sec. 5.8]{Mahler_book14}\cite[Eq. (25)]{Williams15b}
\begin{align}
G_{k|k}[h] & \propto\frac{\delta}{\delta Z_{k}}F[g,h]\bigg|_{g=0}.\label{eq:Update_PGLF_rule}
\end{align}
Applying the product rule \cite[Eq. (31)]{Williams15b}, we obtain

\begin{align}
G_{k|k}[h] & \propto\sum_{W_{0}\uplus\cdots\uplus W_{n_{k|k-1}}=Z_{k}}\frac{\delta}{\delta W_{0}}F^{0}[g,h]\nonumber \\
 & \times\sum_{a\in\mathcal{A}_{k|k-1}}\prod_{i=1}^{n_{k|k-1}}\frac{\delta}{\delta W_{i}}\left(w_{k|k-1}^{i,a^{i}}F^{i,a^{i}}[g,h]\right)\bigg|_{g=0}.\label{eq:UpdatedJointPGFL}
\end{align}
The sum in \eqref{eq:UpdatedJointPGFL} is over all decompositions
of the measurement set $Z$ into $(n_{k|k-1}+1)$ subsets, where one
subset, $W_{0}$, represents measurements which are either false alarms,
or correspond to a target represented by the PPP component (i.e.,
a target which has never been detected so far), and subset $W_{i}$,
$i>0$, represents measurements assigned to the $i$-th Bernoulli.

We develop the required set derivatives over the following lemmas,
starting with the Bernoulli component $F^{i}[g,h]$ in Section \ref{subsec:Bernoulli-update_app},
and then moving on to the update of the PPP, $F^{0}[g,h]$, in Section
\ref{subsec:PPP-update_app}. 

\subsubsection{Bernoulli update\label{subsec:Bernoulli-update_app}}

We calculate the set derivative of $F^{i,a^{i}}[g,h]$. For $W_{i}\neq\emptyset$,
we have 
\begin{align}
\frac{\delta}{\delta W_{i}}F^{i,a^{i}}[g,h] & =r_{k|k-1}^{i,a^{i}}\Big\langle f_{k|k-1}^{i,a^{i}},h\frac{\delta}{\delta W_{i}}G[g|\cdot]\Big\rangle\\
\frac{\delta}{\delta W_{i}}F^{i,a^{i}}[g,h]\bigg|_{g=0} & =r_{k|k-1}^{i,a^{i}}\Big\langle f_{k|k-1}^{i,a^{i}},hf\left(W_{i}|\cdot\right)\Big\rangle\label{eq:updated_PGFL_detection}
\end{align}
where we have applied \eqref{eq:inverse_PGFL}. For $W_{i}=\emptyset$,
the set derivative does not change $F^{i,a^{i}}[g,h]$. 

Equation \eqref{eq:updated_PGFL_detection} and $F^{i,a^{i}}[0,h]$
are the PGFL of a weighted Bernoulli component \cite[Lem. 2]{Williams15b}
with parameters given in the following lemma.
\begin{lem}
\label{lem:MBUpdate} The update of the weighted PGFL component $w_{k|k-1}^{i,a^{i}}G_{k|k-1}^{i,a^{i}}[h]$
(weighted Bernoulli) with measurement set $W_{i}$, i.e., 
\[
w_{k|k}^{i,a^{i},W_{i}}G_{k|k}^{i,a^{i},W_{i}}[h]=\frac{\delta}{\delta W_{i}}\left(w_{k|k-1}^{i,a^{i}}F^{i,a^{i}}[g,h]\right)\big|_{g=0}
\]
is the PGFL of a weighted Bernoulli distribution, i.e., a distribution
of the form \cite[Lem. 2]{Williams15b} 
\begin{equation}
f_{k|k}^{i,a^{i},W_{i}}(X)=w_{k|k}^{i,a^{i},W_{i}}\times\begin{cases}
1-r_{k|k}^{i,a^{i},W_{i}} & X=\emptyset\\
r_{k|k}^{i,a^{i},W_{i}}f_{k|k}^{i,a^{i},W_{i}}(x) & X=\{x\}\\
0 & \left|X\right|>1
\end{cases}
\end{equation}
where for $W_{i}=\emptyset$, 
\begin{align}
w_{k|k}^{i,a^{i},W_{i}} & =w_{k|k-1}^{i,a^{i}}\left[1-r_{k|k-1}^{i,a^{i}}+r_{k|k-1}^{i,a^{i}}l_{k|k}^{i,a^{i},W_{i}}\right]\label{eq:MBClusterMissWeight}\\
l_{k|k}^{i,a^{i},W_{i}} & =\big\langle f_{k|k-1}^{i,a^{i}},f\left(\emptyset|\cdot\right)\big\rangle\\
r_{k|k}^{i,a^{i},W_{i}} & =\frac{r_{k|k-1}^{i,a^{i}}l_{k|k}^{i,a^{i},W_{i}}}{1-r_{k|k-1}^{i,a^{i}}+r_{k|k-1}^{i,a^{i}}l_{k|k}^{i,a^{i},W_{i}}}\label{eq:MBClusterMissPExist}\\
f_{k|k}^{i,a^{i},W_{i}}(x) & =\frac{f_{k|k-1}^{i,a^{i}}(x)f\left(\emptyset|x\right)}{l_{k|k}^{i,a^{i},W_{i}}}.\label{eq:MBClusterMissPDF}
\end{align}
For $|W_{i}|\geq1$, 
\begin{align}
w_{k|k}^{i,a^{i},W_{i}} & =w_{k|k-1}^{i,a^{i}}r_{k|k-1}^{i,a^{i}}l_{k|k}^{i,a^{i},W_{i}}\label{eq:MBClusterUpdateWeight}\\
l_{k|k}^{i,a^{i},W_{i}} & =\big\langle f_{k|k-1}^{i,a^{i}},f\left(W_{i}|\cdot\right)\big\rangle\\
r_{k|k}^{i,a^{i},W_{i}} & =1\label{eq:MBClusterUpdatePExist}\\
f_{k|k}^{i,a^{i},W_{i}}(x) & =\frac{f_{k|k-1}^{i,a^{i}}\left(x\right)f\left(W_{i}|x\right)}{l_{k|k}^{i,a^{i},W_{i}}}.\label{eq:MBClusterUpdatePDF}
\end{align}
\end{lem}
This lemma therefore proves how to update a previous Bernoulli with
a misdetection or a detection hypothesis in Theorem \ref{thm:PMBM_update}. 

\subsubsection{PPP update\label{subsec:PPP-update_app}}

We now turn to calculating the update of the PPP in \eqref{eq:UpdatedJointPGFL}
via the set derivative of $F^{0}[g,h]$, see \eqref{eq:JointPGFLPoissonComponent}.
\begin{lem}
\label{lem:PHDPGFLUpdate} The set derivative of $F^{0}[g,h]$ is
\begin{equation}
\frac{\delta}{\delta W_{0}}F^{0}[g,h]=F^{0}[g,h]\sum_{P\angle W_{0}}\prod_{V\in P}d_{V}[g,h]\label{eq:PHDUpdatePGFLDeriv}
\end{equation}
where 
\begin{equation}
d_{V}[g,h]=\frac{\delta}{\delta V}\left(\big\langle\lambda^{C},g\rangle+\big\langle\lambda_{k|k-1},hG[g|\cdot]\big\rangle\right)\label{eq:PHDUpdatePGFLDerivdVdefn}
\end{equation}
and $\sum_{P\angle W_{0}}$ denotes the sum over all partitions $P$
of $W_{0}$. $\square$
\end{lem}
The proof of Lemma \ref{lem:PHDPGFLUpdate} is in Section \ref{subsec:Derivative_F0_app}.

Following \eqref{eq:UpdatedJointPGFL} , we evaluate the first factor
of \eqref{eq:PHDUpdatePGFLDeriv}, $F^{0}[g,h]$, at $g=0$ to obtain
\begin{align}
F^{0}[0,h] & =\exp\left(\big\langle\lambda_{k|k-1},hf\left(\emptyset|\cdot\right)\big\rangle\right).
\end{align}
This is proportional to the PGFL of a PPP with intensity $\lambda_{k|k}\left(x\right)=f\left(\emptyset|x\right)\lambda_{k|k-1}\left(x\right)$,
which proves \eqref{eq:updated_PPP}.

We now need to compute the set derivatives in \eqref{eq:PHDUpdatePGFLDerivdVdefn},
evaluate them at $g=0$ and compute the corresponding multi-object
densities. For $V=\left\{ v\right\} $ (set with a single element),
we have
\begin{align}
d_{\left\{ v\right\} }[g,h] & =\lambda^{C}\left(v\right)+\big\langle\lambda_{k|k-1},h\frac{\delta}{\delta\left\{ v\right\} }G[g|\cdot]\big\rangle\\
d_{\left\{ v\right\} }[0,h] & =\lambda^{C}\left(v\right)+\big\langle\lambda_{k|k-1},hf\left(\left\{ v\right\} |\cdot\right)\big\rangle\label{eq:Update_PPP_one_detection}
\end{align}
where we have applied the linear rule \cite{Mahler_book14}.

For $\left|V\right|>1$, we have
\begin{align}
d_{V}[g,h] & =\big\langle\lambda_{k|k-1},h\frac{\delta}{\delta V}G[g|\cdot]\big\rangle\label{eq:PGFL_update_PPP_multi_detection}\\
d_{V}[0,h] & =\big\langle\lambda_{k|k-1},hf\left(V|\cdot\right)\big\rangle\label{eq:Update_PPP_multi_detection}
\end{align}
where we have applied that the set derivative of a constant is zero.
This is why the term $\lambda^{C}\left(v\right)$ is not present in
\eqref{eq:PGFL_update_PPP_multi_detection}. 

The following lemma provides the form of the multi-object densities
whose PGFL is $d_{V}[0,h]$ in \eqref{eq:Update_PPP_one_detection}
and \eqref{eq:Update_PPP_multi_detection}. 
\begin{lem}
\label{lem:PPPUpdateComponents} The update of the PGFL of the PPP
prior with measurement subset $V$ 
\[
w_{k|k}^{V}G_{k|k}^{V}[h]=d_{V}[g,h]\Big|_{g=0}
\]
are PGFLs of weighted Bernoulli distributions with the form \cite[Lem. 2]{Williams15b}
\begin{equation}
f_{k|k}^{V}(X)=w_{k|k}^{V}\times\begin{cases}
1-r_{k|k}^{V} & X=\emptyset\\
r_{k|k}^{V}f_{k|k}^{V}(x) & X=\{x\}\\
0 & \left|X\right|>1
\end{cases}
\end{equation}
where
\begin{align}
w_{k|k}^{V} & =\left[\delta_{1}\left[|V|\right]\prod_{z\in V}\lambda^{C}\left(z\right)\right]+l_{k|k}^{V}\label{eq:PoissonClusterUpdateWeight}\\
l_{k|k}^{V} & =\bigg\langle\lambda_{k|k-1},f\left(V|\cdot\right)\bigg\rangle\\
r_{k|k}^{V} & =\frac{l_{k|k}^{V}}{w_{k|k}^{V}}\label{eq:PoissonClusterUpdatePExist}\\
f_{k|k}^{V}(x) & =\frac{f\left(V|x\right)\lambda_{k|k-1}\left(x\right)}{l_{k|k}^{V}}.\quad\square\label{eq:PoissonClusterUpdatePDF}
\end{align}
\end{lem}
Therefore, the PGFL of the updated PPP in \eqref{eq:PHDUpdatePGFLDeriv}
corresponds to the union of a PPP for undetected targets, with intensity
$\lambda_{k|k}\left(x\right)=f\left(\emptyset|x\right)\lambda_{k|k-1}\left(x\right)$
and, a multi-Bernoulli mixture where each term in the mixture is a
partition of $W_{0}$ and each Bernoulli component has a weight and
density provided in Lemma \ref{eq:PHDUpdatePGFLDeriv}. This concludes
the proof of Theorem \ref{thm:PMBM_update}.

It should be noted that to perform the PMBM update we first take all
possible sets $W_{0}\uplus\cdots\uplus W_{n_{k|k-1}}=Z_{k}$, which
represents subsets of $Z_{k}$ associated to the PPP ($W_{0}$) or
the previous Bernoullis ($W_{i}$, $i>0$). Then, we take all possible
partitions $P$ of $W_{0}$, $P\angle W_{0}$, to generate the new
Bernoulli components. A compact way to represent these decompositions
is to take all possible subsets of $Z_{k}$ to generate the new Bernoulli
components and represent the possible data associations to previous
Bernoulli components, as in done in Theorem \ref{thm:PMBM_update}.

\subsection{Set derivative of $F^{0}[g,h]$\label{subsec:Derivative_F0_app}}

We prove Lemma \ref{lem:PHDPGFLUpdate} by induction. The set derivative
of $F^{0}[g,h]$, see \eqref{eq:JointPGFLPoissonComponent}, with
respect to a set with $|W|=1$ is straightforward, as there is a single
partitioning of a one element set. For induction, we assume that the
lemma holds up to a given size $|W|$, and we show that it holds for
$\tilde{W}=W\cup\{z\}$:
\begin{align}
 & \frac{\delta}{\delta\tilde{W}}F^{0}[g,h]\nonumber \\
 & =\frac{\delta}{\delta\{z\}}\frac{\delta}{\delta W}F^{0}[g,h]\\
 & =\frac{\delta}{\delta\{z\}}\left(F^{0}[g,h]\sum_{P\angle W}\prod_{V\in P}d_{V}[g,h]\right)\\
 & =\left(\frac{\delta}{\delta\{z\}}F^{0}[g,h]\right)\sum_{P\angle W}\prod_{V\in P}d_{V}[g,h]\nonumber \\
 & \:+F^{0}[g,h]\sum_{P\angle W}\left(\frac{\delta}{\delta\{z\}}\prod_{V\in P}d_{V}[g,h]\right)\\
 & =F^{0}[g,h]d_{\{z\}}[g,h]\sum_{P\angle W}\prod_{V\in P}d_{V}[g,h]\nonumber \\
 & \,+F^{0}[g,h]\sum_{P\angle W}\sum_{V\in P}\left(\frac{\delta}{\delta\{z\}}d_{V}[g,h]\right)\prod_{V'\in P\backslash\{V\}}d_{V'}[g,h].\label{eq:PHDUpdatePGFLDeriv4}
\end{align}
Each step in the previous derivation results from the product rule
\cite{Mahler_book14}. 

From \eqref{eq:PHDUpdatePGFLDerivdVdefn}, we have $\frac{\partial}{\partial\{z\}}d_{V}[g,h]=d_{V\cup\{z\}}[g,h]$.
In addition, each partitioning of $\tilde{W}$ consists of either
a partitioning of $W$ with an additional single element subset $\{z\}$;
or a partitioning of $W$, adding element $z$ to one of the existing
subsets \cite[App. D.2]{Mahler_book14}. Since the top line in \eqref{eq:PHDUpdatePGFLDeriv4}
handles the former case and the bottom line handles the latter case,
we find that \eqref{eq:PHDUpdatePGFLDeriv4} is equivalent to $F^{0}[g,h]\sum_{P\angle\tilde{W}}\prod_{V\in P}d_{V}[g,h]$,
which proves Lemma \ref{lem:PHDPGFLUpdate}.

\section{\label{sec:IMM_relation_appendix}}

\subsection{Single-target integral}

Given a real-valued function $\pi\left(\cdot\right)$ on $\mathcal{X}=\mathbb{R}^{n_{x}}\uplus\mathcal{X}_{e}$
such that
\begin{align}
\pi\left(x\right) & =\begin{cases}
\pi_{p}\left(x\right) & x\in\mathbb{R}^{n_{x}}\\
\pi_{e}\left(\gamma,\xi,X\right) & x=\left(\gamma,\xi,X\right)\in\mathcal{X}_{e},
\end{cases}
\end{align}
its single target-integral is the sum of the integrals in $\mathbb{R}^{n_{x}}$
and $\mathcal{X}_{e}$ \cite[Sec. 3.5.3]{Mahler_book14}

\begin{align}
\int_{\mathcal{X}}\pi\left(x\right)dx & =\int_{\mathbb{R}^{n_{x}}}\pi_{p}\left(x\right)dx\nonumber \\
 & \:+\int_{\mathbb{S}_{+}^{d}}\int_{\mathbb{R}^{n_{x}}}\int_{\mathbb{R}_{+}}\pi_{e}\left(\gamma,\xi,X\right)d\gamma d\xi dX.
\end{align}

\subsection{Relation to spaces in interacting multiple models}

We explicitly relate the space of coexisting point extended targets,
$\mathcal{X}=\mathbb{R}^{n_{x}}\uplus\mathcal{X}_{e}$, in Section
\ref{sec:Coexisting_point_extended} to spaces used in interacting
multiple models \cite{Mazor98}, which usually include a class variable
to distinguish between different models. Given $x\in\mathbb{R}^{n_{x}}\uplus\mathcal{X}_{e}$,
we know if $x$ represents a point target or an extended target as
$\mathbb{R}^{n_{x}}$ and $\mathcal{X}_{e}$ are disjoint. Therefore,
it is not necessary to extend the single target space with a class
variable to distinguish both types of targets. 

Nevertheless, it is possible to add a class variable $c$ such that
the single target state becomes $y=\left(c,x\right)$, where $c=0$
for point targets and $c=1$ to extended targets. In this case, the
single target space is $\left(\left\{ 0\right\} \times\mathbb{R}^{n_{x}}\right)\uplus\left(\left\{ 1\right\} \times\mathcal{X}_{e}\right)$
and the PMBM filtering recursion remains unchanged. 

\section{\label{sec:Update_rules_appendix}}

For completeness, in this appendix, we provide the (approximate) single
extended target update for factorised GGIW priors \cite{Feldmann11,Granstrom20}.
The resulting expressions are provided in Table \ref{tab:Update-GGIW}.
The update for the parameters of the Gamma distribution is exact due
to the Poisson-Gamma conjugacy.

\begin{table}
\caption{\label{tab:Update-GGIW}Update and marginal likelihood of a GGIW density}

{\fontsize{8}{8}\selectfont

\textbf{Input:} Prior GGIW parameters $\zeta_{+}=\left(\alpha_{+},\beta_{+},\overline{x}_{+},P_{+},v_{+},V_{+}\right)$,
set $W$ of measurements.

\textbf{Output: }$\left(\zeta,\ell\right)=\mathrm{u}_{e}\left(\zeta_{+},W\right)$,
where $\zeta$ are the updated GGIW parameters and $\ell$ the marginal
likelihood evaluated at $W$.

\rule[0.5ex]{1\columnwidth}{1pt}

If $\left|W\right|>0$

\begin{align*}
\zeta & =\begin{cases}
\alpha & =\alpha_{+}+\left|W\right|\\
\beta & =\beta_{+}+1\\
\overline{x} & =\overline{x}_{+}+K\varepsilon\\
P & =P_{+}-KHP_{+}\\
v & =v_{+}+\left|W\right|\\
V & =V_{+}+N+Z
\end{cases}
\end{align*}
where
\begin{align*}
\overline{z} & =\frac{1}{\left|W\right|}\sum_{z\in W}z\\
Z & =\sum_{z\in W}\left(z-\overline{z}\right)\left(z-\overline{z}\right)^{T}\\
\hat{X} & =V_{+}\left(v_{+}-2d-2\right)^{-1}\\
\varepsilon & =\overline{z}-H\overline{x}_{+}\\
S & =HP_{+}H^{T}+\frac{\hat{X}}{\left|W\right|}\\
K & =P_{+}H^{T}S^{-1}\\
N & =\hat{X}^{1/2}S^{-1/2}\varepsilon\varepsilon^{T}S^{-T/2}\hat{X}^{T/2}
\end{align*}
\begin{align*}
\ell & =\left(\pi^{\left|W\right|}\left|W\right|\right)^{-d/2}\frac{\left|V_{+}\right|^{\frac{v_{+}-d-1}{2}}\Gamma_{d}\left(\frac{v-d-1}{2}\right)\left|\hat{X}\right|^{1/2}\Gamma\left(\alpha\right)\left(\beta_{+}\right)^{\alpha_{+}}}{\left|V\right|^{\frac{v-d-1}{2}}\Gamma_{d}\left(\frac{v_{+}-d-1}{2}\right)\left|S\right|^{1/2}\Gamma\left(\alpha_{+}\right)\left(\beta\right)^{\alpha}}.
\end{align*}

If $\left|W\right|=0$

\begin{align*}
\zeta & =\left(\alpha_{+},\beta_{+}+1,\overline{x}_{+},P_{+},v_{+},V_{+}\right)\\
\ell & =\left(\frac{\beta_{+}}{\beta_{+}+1}\right)^{\alpha_{+}}.
\end{align*}

\rule[0.5ex]{1\columnwidth}{1pt}

}

\end{table}

\end{document}